\documentclass[journal,draftclsnofoot,onecolumn,12pt]{IEEEtran}
\usepackage{amsthm,amssymb,graphicx,multirow,amsmath,color,amsfonts}
\usepackage[update,prepend]{epstopdf}
\usepackage[noadjust]{cite}
\usepackage[latin1]{inputenc}
\usepackage{tikz}
\usepackage{bbm} 
\usepackage{pdfpages}
\usepackage{tabulary}
\usepackage{multirow}
\usepackage{comment}
\usepackage{subfigure}
\usepackage{float}

\def\nb0{{\mathbf{0}}}
\def\nb1{{\mathbf{1}}}

\newtheorem{lemma}{Lemma}

\newtheorem{definition}{Definition}

\newtheorem{theorem}{Theorem}

\newtheorem{remark}{Remark}

\allowdisplaybreaks 
\begin{document}
	\title{
		On the Influence of Charging Stations Spatial Distribution on Aerial Wireless Networks
	}
	\author{
		Yujie Qin, Mustafa A. Kishk, {\em Member, IEEE}, and Mohamed-Slim Alouini, {\em Fellow, IEEE}
		\thanks{Yujie Qin, Mustafa A. Kishk, and Mohamed-Slim Alouini are with Computer, Electrical and Mathematical Sciences and Engineering (CEMSE) Division, King Abdullah University of Science and Technology (KAUST), Thuwal, 23955-6900, Saudi Arabia (e-mail: yujie.qin@kaust.edu.sa; mustafa.kishk@kaust.edu.sa; slim.alouini@kaust.edu.sa).} 
		
	}
	
	\maketitle
	\begin{abstract}
		Using drones for cellular coverage enhancement is a recent technology that has shown a great potential in various practical scenarios. However, one of the main challenges that limits the performance of drone-enabled wireless networks is the limited flight time. In particular, due to the limited on-board battery size, the drone needs to frequently interrupt its operation and fly back to a charging station to recharge/replace its battery. In addition, the charging station might be responsible to recharge multiple drones. Given that the charging station has limited capacity, it can only serve a finite number of drones simultaneously. Hence, in order to accurately capture the influence of the battery limitation on the performance, it is required to analyze the dynamics of the time spent by the drones at the charging stations. In this paper, we use tools from queuing theory and stochastic geometry to study the influence of each of the charging stations limited capacity and spatial density on the performance of a drone-enabled wireless network.
		
	\end{abstract}
	
	\section{Introduction}
	
    Unmanned aerial vehicles (UAVs, also known as drones) are expected to play an essential role in potentially enhancing the performance of the next-generation wireless networks \cite{sekander2018multi,mozaffari2019tutorial,8579209,7470933}. Because they can easily function as aerial base stations (BSs) with high relocation flexibility based on dynamic traffic demands, they can be useful in various BS deployment scenarios in both rural and urban areas, such as providing services to remote Internet of Things users \cite{8672604} and improving the quality of service \cite{9244132}. 	UAVs can be deployed in dangerous environments or in natural disasters, such as fires or severe snow storms. In these situations, terrestrial BSs (TBSs) are more likely to be overloaded or heavily damaged, while drones can provide stable connectivity, which makes them a feasible and practical alternative. Moreover, at places where the spatial distributions of active users continuously change with time, UAVs are more flexible than fixed TBSs, since they have the capability to optimize their locations in real-time. Meanwhile, drones can assist TBSs to deliver user hotspots with reliable network coverage and complement existing cellular networks by providing additional capacity \cite{9205314}. In addition, since the altitude of UAVs is adjustable, they are more likely to establish line-of-sight (LoS) links with ground users than TBSs \cite{8675384,8833522}.
	
	Despite the various benefits of UAVs, the UAV's on board energy limitation is one of the main system's bottlenecks. UAVs rely on their internal battery for power supply. Hence, the amount of time they can stay in the air is limited. Consequently, UAVs' offered service is likely to be interrupted, and they are forced to fly back to the charging stations before the battery gets drained. When UAVs recharge, users in UAVs' coverage area experience lower service quality~\cite{9205314}.
	
	Generally, the total energy consumption of UAVs is composed of two parts: communication-related power and propulsion-related  power \cite{mozaffari2019tutorial},  \cite{8675384}. In this work, we consider a scenario where rotary-wing UAVs are deployed to provide wireless coverage to users located at {\em hotspots}~\cite{8663615}.  However, hovering is a power-consuming status, and its corresponding propulsion-related energy highly dominates the communication-related energy. In other words, the reliability, sustainability and feasibility of UAV-assisted networks are greatly restricted by the limited battery lifetime and the recharging methods. In this paper, we use tools from stochastic geometry and queuing theory to study the impact of the capacity of charging stations and their spatial density on the UAV-enabled wireless network's performance. More details on the contributions of this paper are provided in Sec~\ref{sec:cont}.
	
	\subsection{Related Work}
	Literature related to this work can be categorized into: (i) flight duration enhancement using energy harvesting, (ii) innovative system architectures to extend UAV's endurance, and (iii) stochastic geometry-based frameworks for UAV wireless networks. A brief discussion on related works in each of these categories is discussed in the following lines.
	
    {\em Energy Harvesting UAVs.}  One potential solution to enhance the flight duration of the drones in a UAV-enabled wireless networks is to exploit the advances in the energy harvesting technology. In urban communication environments, authors in~\cite{8501925} studied a UAV-based relaying system which harvests energy from ground BSs. For that setup, they derived the lower bound for outage probability considering various UAV altitudes.  Authors in \cite{8494684}  used the radio frequency (RF) energy harvesting technology to enhance the lifetime of the UAV battery. To maximize the throughput, dirty paper coding scheme was considered, as well as uplink beamforming and downlink power control. Energy harvesting from solar or wind resources was analyzed in \cite{9060991}. Based on their statistic model, authors derived the probability density function (PDF), cumulative density function (CDF) of the amount of energy harvested from the above renewable energy resources, and outage probability expressions.
    Authors in \cite{mekikis2019breaking} propose the use of solar-powered charging stations to satisfy the energy need of UAVs, and use matching theory to solve the allocation problem. In \cite{8413129}, authors improved energy efficiency of UAVs by route planning based on  dynamic programming.

{\em Alternative System Architectures.} The system architecture of the UAV-enabled wireless network can be modified for the sake of a longer flight time~\cite{8648453}. Firstly, the influence of frequently interrupting and revisiting the charging stations was studied in~\cite{9153823} with emphasis on signal-to-noise-ratio (SNR) and the assumption that the charging stations have infinite capacity. Authors in~\cite{9205314,kishk20193d,bushnaq2020cellular}, studied a system where the UAV is physically connected to a ground station through a tether. This tether provides the UAV with a stable power supply and a reliable data link. However, the tether restricts the mobility of the UAV. Authors in~\cite{8866716,8403572,8648007,jaafar2020dynamics} studied a system where laser beam directors (LBDs) are located on the ground and directing their laser beams towards UAVs to provide them with the required energy. Similar to the tethered UAV, laser-powered UAV still needs to be relatively close to the LBD in order to receive enough energy through the laser beam and to ensure LoS.

{\em Stochastic Geometry-based Literature.} Stochastic geometry is a strong mathematical tool that enables characterizing the statistics of various large-scale wireless networks~\cite{7733098,6524460}. It was used in~\cite{8833522} to study a heterogeneous network composed of terrestrial and aerial BSs with both spatially distributed according to two independent Poisson point processes (PPPs). For that setup, after accurately characterizing the Laplace transform of the interference coming from both aerial and terrestrial BSs, downlink coverage probability and average data rate were derived. Authors in \cite{8713514} derived the coverage probability for a UAV-enabled cellular network where UAVs are deployed at the centers of user hotspots. The locations of the hotspot centers are modeled as a PPP while the locations of the users are modeled using Matern cluster process (MCP)~\cite{7809177}. Binomial point process was also used to model the locations of a given number of UAVs deployed in a finite area while assuming static locations in~\cite{7967745}, and dynamic locations in~\cite{8681266}. Authors in~\cite{8017572} considered a setup where a single UAV provides wireless coverage to ground users with the assistance of randomly-located ground relayes.

    While the existing literature focus on enhancing UAV's performance by using energy harvesting, improving system architectures and stochastic geometry-based tools, there is no work to analyze the impact of limited charging resources.

	\subsection{Contribution}\label{sec:cont}
In this paper, our objective is to study the influence of the spatial distribution of the UAV-charging stations and their capacity (maximum number of UAVs that can be recharged simultaneously) on the coverage probability of a UAV-enabled wireless network. Hence, we consider a setup where hotspot centers and charging/swapping stations are spatially distributed according to two independent PPPs. More detailed discussion on this paper's main contributions is provided next.

{\em Novel Framework and Performance Metrics}. We introduce a novel performance metric, the UAV's availability probability, which is defined as the probability that the UAV has enough energy in its battery to hover and provide cellular service. We provide a mathematical definition for this probability as a function of the battery size, the power consumption, the time required for recharging/swapping, the distance to the nearest charging station, and the time spent at the charging station's queue.	 Next, given that the last two parameters are random variables, we compute the average value of the availability probability, using tools from stochastic geometry and queuing theory.

{\em Coverage Probability}. While the coverage probability of a UAV-enabled wireless network is a well-established result in literature, we revisit its definition by incorporating the UAV's availability probability into the coverage probability definition. Hence, our framework leads to more accurate expressions for the coverage probability that captures the influence of various system parameters that are typically ignored in literature, such as the battery size and the capacity and spatial density of the charging stations.

{\em System-Level Insights}. Using the reformulated expressions for the coverage probability, our numerical results reveal various useful system level insights. We show that slightly increasing the charging station's capacity significantly reduces the density of charging stations required to achieve a specific level of coverage probability. Furthermore, we show that increasing the charging station's capacity is only beneficial upto a specific value, afterwards, the coverage probability becomes constant.

	\section{System Model}
	
\begin{figure}[ht]
		\centering
		\subfigure[]{\includegraphics[width=0.47\columnwidth]{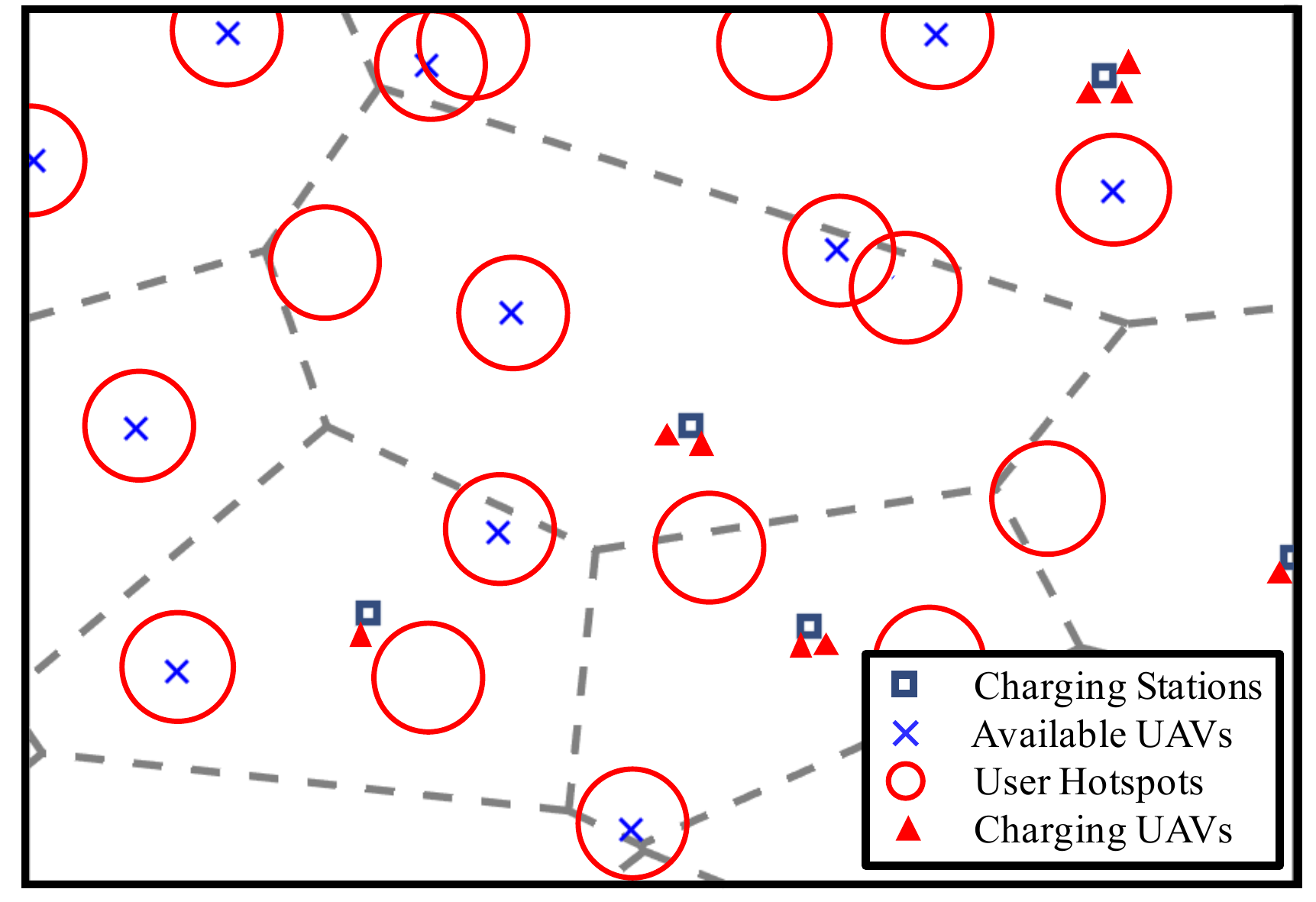}}
		\subfigure[]{\includegraphics[width=0.47\columnwidth]{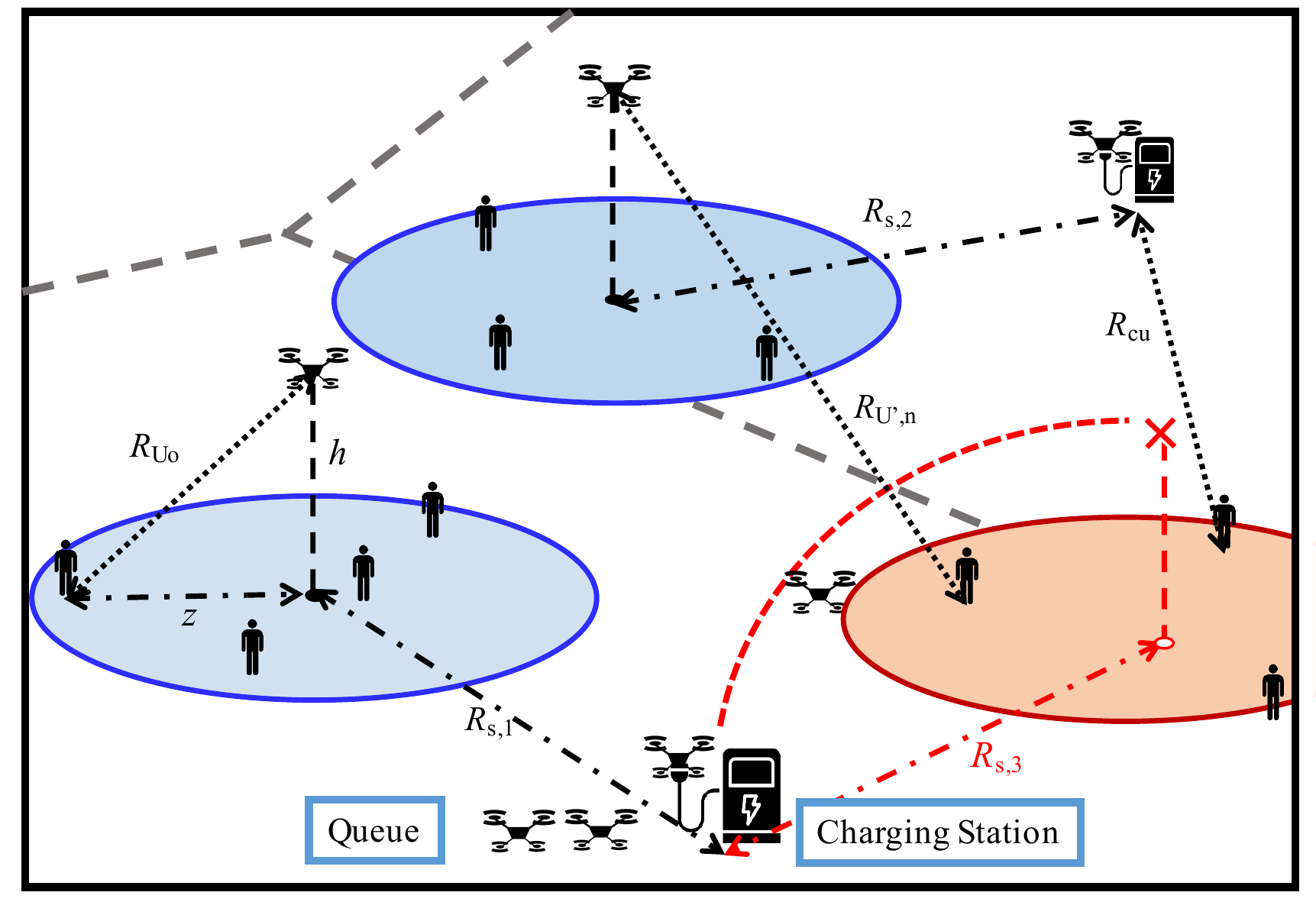}}
		\caption{Illustration of the system model.}
		\label{fig_sys_1}
	\end{figure}

	We consider a cellular network where UAVs and charging stations are spatially-distributed according to two independent homogeneous PPPs, $\Phi_{\rm u}$ and $\Phi_{\rm c}$, with densities $\lambda_{\rm u}$ and $\lambda_{\rm c}$, respectively. As mentioned in \cite{8713514}, MCP is widely used in modeling user distribution, we adapt this model in our system. In MCP, the clusters are modeled as disks with radii $r_c$ whose centers are modeled as a PPP while the users at each cluster are uniformly distributed within the disk. The UAVs are assumed to hover at a fixed altitude of $h$ above each hotspot center. Given that each UAV flies back to the nearest charging station before running out of energy, the association regions of the UAVs with the charging stations form a Poisson-Voronoi (PV) tessellation, as depicted in Fig. \ref{fig_sys_1} (a). Using Slivnyak's theorem \cite{haenggi2012stochastic}, without loss of generality, we perform our analysis in the rest of the paper at a typical UAV located at the origin and the typical PV cell that contains the origin.
	
	\begin{table*}[ht]\caption{Table of Notations}
		\centering
		\begin{center}
			\resizebox{\textwidth}{!}{
				\renewcommand{\arraystretch}{1}
				\begin{tabular}{ {c} | {c} }
					\hline
					\hline
					\textbf{Notation} & \textbf{Description} \\ \hline
					$\Phi_{\rm c}$, $\Phi_{\rm c,a}$; $\lambda_{\rm c}$, $\lambda_{\rm c}^{'}$ & PPP of charging stations, PPP of active charging stations; density of the charging stations, density of active charging stations\\ \hline
					$\Phi_{\rm u}$, $\Phi_{\rm u^{'}}$; $\lambda_{\rm u}$, $\lambda_{\rm u}^{'}$ & PPP  of  UAVs, PPP of available UAVs; density of  UAVs, density of available UAVs\\ \hline
					$\Phi_{\rm u_{o}}$; $\Phi_{\rm u^{'}_{l}}$; $\Phi_{\rm u^{'}_{n}}$ & Location of the typical UAV, available LoS UAVs, available NLoS UAVs, respectively \\ \hline
					$R_{\rm s}$, $R_{\rm c}$ & Horizontal distances between the typical UAV and the typical charging charging station, and the nearest active charging station, respectively \\ \hline
					$c$; $N$; $Ratio$ & Charging station capacity; the number of UAVs in the typical PV cell; refers to $\lambda_{\rm u}/\lambda_{\rm c}$ \\ \hline
					$S_i$, $S_{(i,j)}$; $p_x$, $P_i$ & Waiting time states, substates; probability that the charging station holds $x$ UAVs, and that it stays in state $S_i$ \\ \hline
					$h$; $a_{\rm ave}$ &  UAV altitude; average acceleration while landing/taking off \\ \hline
					$V_{\rm max}$, $V$ & Maximum velocity while landing or taking off, UAV's velocity during traveling \\ \hline
					$T_{\rm se}$, $T_{\rm se,E}$ & Service time, expectation of service time \\\hline
					$T_{\rm w}(i)$ & Waiting time in state $i$ \\\hline
					$T_{\rm tra}$, $T_{\rm tra,E}$ & Time required to travel to or from the nearest charging station, expectation of traveling time\\\hline
					$T_{\rm ch}$, $T_{\rm land}$ & Time used in recharging and landing or taking off, respectively \\\hline
					$E_l$, $E_t$ & Energy consumed in landing or taking off and in  traveling, respectively\\ \hline
					$C_{\rm R_s}$, $C_{\rm R_c}$ & Typical charging station which is the nearest to the origin and the nearest active charging station (excluding $C_{\rm R_s}$)\\ \hline
					$P_{\rm a}$, $P_{\rm C,a}$, $P_{\rm Crs,a}$ & UAV's availability probability, activity probability of charging stations, and $C_{\rm R_s}$, respectively\\ \hline
					$P_{\rm m}$, $P_{\rm s}$ & Power consumed during traveling and service, respectively \\\hline
					$R_{\rm U_{o}}$, $R_{\rm U^{'},{l}}$, $R_{\rm U^{'},{n}}$ & Distances between the typical user and the typical UAVs, nearest available LoS UAV, and nearest available NLoS UAV, respectively \\ \hline
					$R_{\rm su}$, $R_{\rm cu}$ & Distances between the typical user and $C_{\rm R_s}$ and $C_{\rm R_c}$, respectively \\ \hline
					$\mathcal{A}_{\rm LoS}$, $\mathcal{A}_{\rm NLoS}$ & Probability of associating with nearest LoS UAVs and NLoS UAVs, respectively \\\hline
					$\mathcal{A}_{\rm Cs}$, $\mathcal{A}_{\rm Cc}$ & Probability of associating with  $C_{\rm R_s}$ and $C_{\rm R_c}$, respectively \\\hline\hline
			\end{tabular}}
		\end{center}
	\end{table*}
	
	\subsection{UAV's Availability} 
	
	We consider a scenario where UAVs are supposed to fly back to the nearest charging station as soon as the battery charge drops to a specific level, which is function of the amount of energy needed for traveling towards the charging station, and hence, the distance to the charging station. The capacity of each charging stations $c$ is finite, which means they can only charge\footnote{     In this paper, we use the term "charging time" to refer to either (i) battery swapping or (ii) battery charging.} $c$ UAVs simultaneously. Hence, the UAVs may wait in a queue, and the waiting time $T_{\rm w}$ depends on the length of the queue. For the queue analysis, we consider a discrete-time series with the time slot of length $T_{\rm ch}$, during which at most $c$ UAVs can be charged simultaneously. To enable analytical tractability, we assume that the charging process starts at the beginning of each time slot and the charging UAV leaves at the end of the slot. For a given number of UAVs $N$ in the typical PV cell, we assume that each of the rest of the UAVs, aside from the typical UAV located at the origin, have the following probability of being at the typical charging station
	\begin{align}\label{eq1}
	P_{\rm ch}(i)=\frac{T_{\rm ch}+T_{\rm w}(i)}{T_{\rm ch}+T_{\rm w}(i)+2T_{\rm land}+2T_{\rm tra,E}+T_{\rm se,E}},
		\end{align} 
		where $T_{\rm ch}$ is the time required for charging, $T_{\rm tra,E}$ and $T_{\rm se,E}$ are the average values of the required time to travel to and from the nearest charging station and the time spent at the hotspot center to provide service, respectively, and $T_{\rm land}$ is the time spent during landing or taking off. Accordingly, $T_{\rm tra,E}$, $T_{\rm se,E}$, $T_{\rm land}$ and $T_{\rm w}$ can be formally defined as follows
	\begin{align}
	T_{\rm se,E}&=\frac{B_{\rm max}-2P_{\rm m}\frac{\mathbb{E}[R_{\rm s}]}{V}-2E_l}{P_{\rm s}},\label{eq_Tse_E}\\
	T_{\rm tra,E}&=\frac{\mathbb{E}[R_{\rm s}]}{V},\\
	T_{\rm land}&=2\sqrt{\frac{2h}{a_{\rm ave}}},\label{eq_Tland}\\
	T_{\rm w}(i)&=i\times T_{\rm ch},\label{eq_Tw}
	\end{align}
	where ${B}_{\rm max}$ is the UAV battery size, $R_{\rm s}$ is the distance between a UAV and its nearest charging station, $P_{\rm m}$ denotes the power consumption during traveling, $V$ is the UAV's velocity while traveling, $P_{\rm s}$ is the power consumption during hovering at the hotspot center, which includes both the propulsion power and the total communication power, $a_{\rm ave}$ is the average acceleration while landing and taking off, $E_l$ is the corresponding energy consumption, $\mathbb{E}[\cdot]$ denotes expectation operator, and $T_{\rm w}(i)$ is the waiting time. 
	
	The value of $T_{\rm w}(i)$ is a function of the state of the queue at the charging station $i$, which is explained in the below definition.
	\begin{definition}[Waiting Time State]
		We define different states $S_{i}$ and substates $S_{(i,j)}$, in which $i$ reflects the waiting time $T_{\rm w}(i) = i\times T_{\rm ch}$, $j$ denotes that there are $ci+j$ UAVs at the charging station, and $j<c$ holds for all scenarios. Let $P_{i}(t)$ denote the probability the the charging stations staying is in state $S_i$ at time $t$, in steady state we have
		\begin{align}
		\lim_{t\to \infty}P_{i}(t) = P_{i}\nonumber.
		\end{align}
	\end{definition}	
	When the charging station is at the state $S_{(0,j)}$, at most $c-j$ UAVs that arrive during a given time slot will finish charging before the beginning of the next time slot.
	
	As mentioned, UAVs are available and can provide service to users when they are not traveling to charging stations or waiting in the queue.
	\begin{definition}[Availability Probability]
	      We define the event $\mathcal{A}$ that indicates the availability of the typical UAV, which denotes that the UAV is hovering and provides service. Conditioned on  $N$ UAVs in the typical PV cell, the availability probability, which is a fraction of time, of the UAV is
		\begin{align}
		\mathbb{P}(\rm \mathcal{A} | N)&=\sum_{i}P_{i}\mathbb{E}_{\rm \Phi_{c}}\bigg[\frac{T_{\rm se}(x)}{T_{\rm se}(x)+T_{\rm ch}+T_{\rm w}(i)+2T_{\rm tra}(x)+2T_{\rm land}}\bigg],\label{eq_P_aN}
		\end{align}
		where,
		\begin{align}
		T_{\rm tra}(x)&=\frac{R_{\rm s}(x)}{V}\label{eq_Ttra},\\
		T_{\rm se}(x)&=\frac{B_{\rm max}-2P_{\rm m}\frac{R_{\rm s}(x)}{V}-2E_l}{P_{\rm s}}.\label{eq_Tse}
		\end{align}
		Hence, the uncoditioned availability probability is 
		\begin{align}
		P_{\rm a}&=\mathbb{E}_{\rm N}[\mathbb{P}(\rm \mathcal{A} | N)],\label{def_eq_Pa}
		\end{align}
		where $x$ annotates the typical UAV's location. 
	\end{definition}
	
	In order to enhance the performance of the network and reduce the influence of the frequent recharging process, we consider a scenario where the UAV can reactivate itself and provide service as soon as at reaches the charging station. In that case, the charging stations can behave like a TBS if at least one UAV is recharging.
	\begin{definition}[Active Charging Station]
		An active charging station is a charging station that is occupied by at least one UAV. The point process modeling the locations of active charging stations is denoted as $\Phi_{\rm c,a}$, with density $\lambda_{\rm c}^{'}=\lambda_{\rm c}P_{\rm C,a}$, in which
		\begin{align}
		P_{\rm C,a} &= 1-P_{\rm S_{(0,0)}},\nonumber\\
		P_{\rm S_{(0,0)}}&=\sum_{n=0}^{\infty}P_{\rm (S_{(0,0)}|N)}\mathbb{P}(N=n)\nonumber,
		\end{align}
		where $P_{\rm S_{(0,0)}}$ is the probability the the queuing system staying in state $S_{\rm (0,0)}$. We refer to $P_{\rm C,a}$ as the activity probability in the rest of the paper.
		
When the typical UAV is not available, the activity probability of the typical charging station $C_{\rm R_s}$ is different from $P_{\rm C,a}$ and can be computed as follows   		
		\begin{align}
		P_{\rm Crs,a} = 1-P_{\rm S_{(0,0)}}(1-P_r)\nonumber,
		\end{align}
		 where $P_r$ is the probability that the typical UAV is either charging or waiting at the queue of the typical charging station, given that the typical UAV is unavailable, which can be computed as follows
		\begin{align}
		P_r = \sum_{i=0}P_{i}\mathbb{E}_{\rm \Phi_{c}}\bigg[\frac{T_{\rm w}(i)+T_{\rm ch}}{2T_{\rm land}+2T_{\rm tra}(x)+T_{\rm w}(i)+T_{\rm ch}}\bigg]\nonumber.
		\end{align}
	\end{definition} 
	
To analyze the coverage probability of this setup, it is important to characterize the distance distribution between the cluster center and (i) the typical charging station $C_{\rm R_s}$, and (ii) the nearest active charging station in the point process 	$\Phi_{\rm c,a}\backslash C_{\rm R_s}$.
	
	
	\subsection{Power Consumption}
	
	We consider the UAV's power consumption composed of three parts: (i) service-related power $P_{\rm s}$, including hovering and communication-related power, (ii) traveling power $P_{\rm m}$, which denotes the power consumed in traveling to/from the nearest charging station  through the horizontal distance $R_{\rm s}$, and (iii) landing and taking off energy $E_l$, which owes to the difference in height between the charging stations and UAV's altitude.
	
	Based on \cite{8663615}, ${P}_{\rm m}$ is a function of the UAV's velocity $V$ and given by
	\begin{align}
	P_{\rm m}=P_{\rm 0}\left(1+\frac{3V^2}{U_{\rm tip}^{2}}\right)+\frac{P_{\rm i}v_{\rm 0}}{V}+\frac{1}{2}d_{\rm 0}\rho s AV^{3},\nonumber
	\end{align}
	where $P_{\rm 0}$ and $P_{\rm i}$ present the blade profile power and induced power, $U_{\rm tip}$ is the tip speed of the rotor blade, $v_{\rm 0}$ is the mean rotor induced velocity in hover, $\rho$ is the air density, $A$ is the rotor disc area, $d_{\rm 0}$ is fuselage drag ratio, and $s$ is rotor solidity. Therefore, the energy consumed during traveling to or from the charging station is 
	\begin{align}
	E_t&=\frac{R_{\rm s}(x)}{V}P_{\rm m} \nonumber\\
	&=\frac{R_{\rm s}(x)}{V}\left(P_{\rm 0}\left(1+\frac{3V^2}{U_{\rm tip}^{2}}\right)+\frac{P_{\rm i}v_{\rm 0}}{V}+\frac{1}{2}d_{\rm 0}\rho s AV^{3}\right).\nonumber
	\end{align}
	We assume that the optimal value of $V$ that minimizes $E_{t}$ is used. Similarly, the energy consumed during during landing/taking off is
	\begin{align}
	E_l=\int_{0}^{\sqrt{\frac{2h}{a_{\rm ave}}}}P_{m}(a_{\rm ave}t)t{\rm d}t+\int_{0}^{\sqrt{\frac{2h}{a_{\rm ave}}}}P_{m}(V_{\rm max}-a_{\rm ave}t)t{\rm d}t,\nonumber
	\end{align}
	in which,
	\begin{align}
	V_{\rm max}=\sqrt{2ha_{\rm ave}},\nonumber
	\end{align}
	where $a_{\rm ave}$ denotes the average acceleration while landing or taking off.
	
	\subsection{User Association}
	\label{user_association}
Without loss of generality, we focus on a typical user randomly selected from the typical hotspot centered at the origin. The user associates with the UAV deployed at its hotspot center if it is available. The set $\Phi_{\rm u_{o}}$ is composed of only one point, which is the location of the typical UAV, when it is available, otherwise, $\Phi_{\rm u_{o}}=\emptyset$. If it is unavailable (for charging purposes), the user associates with the UAV in $\Phi_{\rm u^{'}}$ (which presents the locations of all available UAVs) or the active charging station that provides the largest average received power, as depicted in Fig.~\ref{fig_sys_1} (b). The point process $\Phi_{\rm u^{'}}$ is constructed by independently thinning $\Phi_{\rm u}$ with the probability $P_{\rm a}$. Hence, the density of $\Phi_{\rm u^{'}}$ is $\lambda_{u}^{'}=P_{\rm a}\lambda_{\rm u}$.
	
	
	When the typical user associates with a UAV, the received power is
	\begin{align}
	p_{\rm u}&=\left\{ 
	\begin{aligned}
	p_{\rm l}=\eta_{\rm l}\rho_{\rm u}G_{\rm l}R_{\rm u}^{-\alpha_{\rm l}},  & \quad \text{\rm in case of LoS},\\
	p_{\rm n}=\eta_{\rm n}\rho_{\rm u}G_{\rm n}R_{\rm u}^{-\alpha_{\rm n}},  & \quad \text{\rm in case of NLoS},\\
	\end{aligned} \right.\nonumber
	\end{align}
	where $\rho_{\rm u}$ is the transmission power of the UAVs, $R_{\rm u}$ denotes the distance between the typical user and the serving UAV, $\alpha_{\rm l}$ and $\alpha_{\rm n}$ present the path-loss exponent, $G_{\rm l}$ and $G_{\rm n}$ are the fading gains that follow gamma distribution with shape and scale parameters $(m_{\rm l},\frac{1}{m_{\rm l}})$ and $(m_{\rm n},\frac{1}{m_{\rm n}})$, $\eta_{\rm l}$ and $\eta_{\rm n}$ denote the mean additional losses for LoS and NLoS transmissions, respectively. The probability of establishing an LoS link between the typical user and a UAV at distance $R_{\rm u}$ is given in~\cite{6863654} as
	\begin{align}
	P_{\rm l}(R_{\rm u}) & =  \frac{1}{1+A \exp\bigg(-B\bigg(\frac{180}{\pi}\arctan\bigg(\frac{h}{\sqrt{R_{\rm u}^2-h^2}}\bigg)-A\bigg)\bigg)} ,\label{pl_pn}
	\end{align}
	where $A$ and $B$ are two variables that depend on the type of the environment (e.g., urban, dense urban, and highrise urban), and $h$ is the altitude of the UAV. Consequently, the probability of NLoS is $P_{\rm n}(R_{\rm u})=1-P_{\rm l}(R_{\rm u})$.
	
	When the user associates with an active charging station, the received power is
	\begin{align}
	p_{\rm c} &= \rho_{\rm u} H R_{\rm \{su,cu\}}^{-\alpha_{\rm t}},\nonumber
	\end{align}
	in which $R_{\rm \{su,cu\}}$ denotes the distances between the user and $C_{\rm  R_s}$ and $C_{\rm R_c}$ (which are the typical charging station and the nearest active charging station), respectively, $H$ is the fading gain that follows exponential distribution with unity mean, and $\alpha_{\rm t}$ presents the path-loss exponent.
	
	The typical user is successfully served if the SINR of the serving link is above a predefined threshold. We refer to the probability of the SINR greater than this threshold as the coverage probability.
	\begin{definition}[Coverage Probability] 
		\label{def_cov}
		The total coverage probability is defined as 
		\begin{align}\label{cov:eq}
		P_{\rm cov} &= P_{\rm a}P_{\rm cov,U_o}+(1-P_{\rm a})P_{\rm cov,\bar{U}_o},
		\end{align}	
		in which,
		\begin{align}
		P_{\rm cov,{\{U_o,\bar{U}_o\}}} = \mathbb{P}\left({\rm SINR}_{\{U_o,\bar{U}_o\}}\geq\theta\right) ,\nonumber
		\end{align}
		where $P_{\rm cov,{U_o}}$ and $P_{\rm cov,{\bar{U}_o}}$ are the coverage probabilities when the typical UAV is available and unavailable, respectively. Let $\Phi_{\rm u^{'}_l}$ and $\Phi_{\rm u^{'}_n}$ be subsets of $\Phi_{\rm u^{'}}$, which denote the locations of LoS UAVs and NLoS UAVs, respectively.  Conditioning on the serving UAV (or active charging station) located at $u_s$, the aggregate interference is defined as
		\begin{align}
		I &= \sum_{N_i\in\Phi_{\rm u^{'}_n}/u_{s}}\eta_{\rm n}\rho_{\rm u}G_{\rm n}D_{\rm N_i}^{-\alpha_{\rm n}}+\sum_{L_j\in\Phi_{\rm u^{'}_l}/u_{s}}\eta_{\rm l}\rho_{\rm u}G_{\rm l}D_{\rm L_j}^{-\alpha_{\rm l}}+\sum_{C_k\in\Phi_{\rm c,a}\cup\{C_{\rm R_s}\}/u_{s}}\rho_{\rm u}HD_{\rm C_k}^{-\alpha_{\rm t}},\nonumber
		\end{align} 
		in which $D_{\rm N_i}$, $D_{\rm L_j}$ and $D_{\rm C_k}$ are the distances between the typical user and the interfering NLoS, LoS UAVs, and active charging stations, respectively.
	\end{definition}
	
	
	\section{Availability Probability}
	To capture the waiting time of the reference UAV in the charging station, we first derive the probability distribution of the number of UAVs lying in the typical PV cell.
	
	The association region of a charging station is the region of the Euclidean plane in which all UAVs are served by the corresponding charging station. The association cells form the PV cells generated by $\Phi_{\rm c}$. Moreover, the typical UAV is more likely to lie in a larger association cell than in a smaller one. In other words, the area of the typical  PV cell is biased. 
	
	\begin{lemma}[Number of UAVs Inside a Biased Area]
		\label{lem_pv_n_pmf}
		The PMF of the number of UAVs falling in the typical PV cell is given by
		\begin{align}
		\label{eq_lem_pv_n_pmf}
		P(N=n)=\frac{\Gamma(a+n+1)}{\Gamma(a)}\frac{b^a}{n!}\frac{\left(\frac{\lambda_{\rm u}}{\lambda_{\rm c}}\right)^{n}}{\left(b+\frac{\lambda_{\rm u}}{\lambda_{\rm c}}\right)^{a+n+1}},
		\end{align}
		in which $a$ and $b$ are two fitting parameters for the area of PV cells, and $\Gamma(\cdot)$ denotes the Gamma function. 
		\begin{IEEEproof}
			See Appendix \ref{app_pv_n_pmf}.
		\end{IEEEproof}
	\end{lemma}
	In the rest of the paper, we refer to the ratio $\frac{\lambda_{\rm u}}{\lambda_{\rm c}}$ as $Ratio$. 
	Having characterized the distribution of the number of UAVs inside the typical PV cell, we now perform  analysis at waiting time states. 
	
	We consider a scenario where the typical cell contains  $N$ UAVs, including the typical one. At the beginning of a new time interval, say $t_{\rm ch}(1)$, the system starts at state $S_{i_1}$ and substate $S_{(i_1,j_1)}$. If so, there are $ci_1+j_1$ UAVs in that cell and the waiting time is $T_{\rm w}(i_1)=i_1\times T_{\rm ch}$. Let $m$ denotes the number of UAVs that are not in the charging station, which equals to $N-ci_1-j_1$. If $k$  UAVs come to the charging station during this time slot, then the system  transfers to a new state $S_{i_2}$ and  substate $S_{(i_2,j_2)}$ at the beginning of a the next time slot $t_{\rm ch}(2)$. The number of UAVs arriving to the charging station during $(t_{\rm ch}(1),t_{\rm ch}(2)]$ is modeled as a Binomial random variable with an arrival rate $P_{\rm ch}(i_1)$.
	
	\begin{lemma}[Number of UAVs at charging station]
		Let $p_{n_1}(t_{\rm ch}(1))$ be the probability of the charging station holds $n_1$ UAVs at time $t_{\rm ch}(1)$, in which $n_1=ci_1+j_1$. Consequently, $n_2$ UAVs present at time $t_{\rm ch}(2)$ either arrived during the $(t_{\rm ch}(1),t_{\rm ch}(2)]$ or were already waiting  at time $t_{\rm ch}(1)$. 
		The probability of having $n_2$ UAVs at the charging station at the beginning of a new time slot $t_{\rm ch}(2)$ is
		\begin{align}
		p_{n_2}(t_{\rm ch}(2)) =& \left\{ 
		\begin{aligned}
		&\sum_{n_1=0}^{c} p_{n_1}(t_{\rm ch}(1))\sum_{k_1=0}^{c-n_1} \binom{m}{k_1} P_{\rm ch}^{k_1}(i_1)(1-P_{\rm ch}(i_1))^{m-k_1}, \quad n_2=0,\\
		&\sum_{n_1=0}^{\min(N-c,c+n_2)} p_{n_1}(t_{\rm ch}(1)) \binom{m}{k} P_{\rm ch}^{k}(i_1)(1-P_{\rm ch}(i_1))^{m-k}, \quad 0<n_2<N-c,\\
		\end{aligned}\right.\nonumber
		\end{align}
		in which, $m=N-ci_1-j_1$, $k=n_2+c-n_1$ and $P_{\rm ch}(i)$ is given in (\ref{eq1}). The stationary distribution of $p_{n_2}$ can be derived as follows
		\begin{align}
		p_{n_2}=& \left\{ 
		\begin{aligned}
		&\sum_{n_1=0}^{c} p_{n_1}\sum_{k_1=0}^{c-n_1} \binom{m}{k_1} P_{\rm ch}^{k_1}(i_1)(1-P_{\rm ch}(i_1))^{m-k_1}, \quad n_2=0,\\
		&\sum_{n_1=0}^{\min(N-c,c+n_2)} p_{n_1}\binom{m}{k} P_{\rm ch}^{k}(i_1)(1-P_{\rm ch}(i_1))^{m-k}, \quad 0<n_2<N-c.\\
		\end{aligned}\right.\nonumber
		\end{align}
		Solving the above system of equations, along with $\sum_{n_1=0}^{N} p_{n_1} = 1$, enables computing the values of $p_{n_i}$.
		\begin{IEEEproof}
			The relationships between $p_{(0,n_1,...N)}$ can be simply considered as the difference between new comers and those finished charging during $(t_{\rm ch}(1),t_{\rm ch}(2)]$. The corresponding arrival process is modeled by the binomial distribution  $B(k,P_{ch}(i))$. That is, $(t_{\rm ch}(1),t_{\rm ch}(2)]$ is an arbitrary time interval  of length $T_{\rm ch}$, during which $k$ new arrivals will take place with probability $\binom{m}{k}P_{\rm ch}^{k}(i_1)(1-P_{\rm ch}(i_1))^{(m-k)}$. 
		\end{IEEEproof}
	\end{lemma}
	
	Now we proceed to present the probability of waiting time state.
	\begin{lemma}[Steady State]
		The steady state probability $P_i$ is given by
		\begin{align}
		P_{i} = \sum_{n=ci}^{c(i+1)-1}p_{n}.\nonumber
		\end{align}
		\begin{IEEEproof}
			Observe that $p_{n}$ indicates the probability of $n$ UAVs  in the charging station and $p_{ci}$ to $p_{c(i+1)-1} $ reflect the probability of state $S_{i}$. Summing $p_{ci}$ to $p_{c(i+1)-1}$ directly completes the proof.
		\end{IEEEproof}
	\end{lemma}
	
	As stated earlier, $T_{\rm w}(i)$ has an impact on availability probability and varies from one cell to the other. Conditioned on a typical PV cell, which contains the typical UAV located at the origin, we now can derive the conditioned availability probability.
	\begin{lemma}[Conditioned Availability Probability] \label{lem_cond_Pa}
		For the analysis that follows, let $I_{\rm max}$ be the last state that has the longest waiting time, in which $I_{\rm max} = \lfloor\frac{N}{c}\rfloor$.	Given the value of $N$, the availability probability can be written as
		\begin{align}
		\label{eq_lem_cond_Pa}
		P_{(\rm a|N)}&=\sum_{i=0}^{I_{\rm max}-1}P_{i}\int_{a_5}^{\frac{a_1}{a_3(i)}}1-\exp\bigg(-\lambda_{\rm c}\pi \bigg(\frac{-a_1+a_3(i)y}{-a_2-a_4y}\bigg)^2\bigg){\rm d}y,
		\end{align}
		in which,
		\begin{align}
		a_1&=V(B_{\rm max}-2E_l),\nonumber\\
		a_2&=2P_{\rm m},\nonumber\\
		a_3(i)&=V\bigg(B_{\rm max}-2E_l+P_{\rm s}T_{\rm ch}(1+i)+4P_{\rm s}\sqrt{\frac{2h}{a_{\rm ave}}}\bigg),\nonumber\\
		a_4&=2(P_{\rm s}-P_{\rm m}),\nonumber\\
		a_5 &=  \frac{2P_{\rm m}a_1-a_2V(B_{\rm max}-2E_l)}{2P_{\rm m}a_3(i)+a_4V(B_{\rm max}-2E_l)}.\nonumber
		\end{align}
		\begin{IEEEproof}
			See Appendix \ref{app_cond_Pa}
		\end{IEEEproof}
	\end{lemma}
	
	In the following theorem, we derive the availability probability.
	\begin{theorem}[Availability Probability]
		The availability probability of the UAV  is
		\begin{align}
		P_{\rm a}&=\sum_{n=0}^{\infty}P_{(\rm a|N) }\mathbb{P}(N=n) \nonumber\\
		&=\sum_{n=0}^{\infty}P_{(\rm a|N)}\frac{\Gamma(a+n+1)}{\Gamma(a)}\frac{b^a}{n!}\frac{\lambda_{\rm c}^{a+1}\lambda_{\rm u}^{n}}{(b\lambda_{\rm c}+\lambda_{\rm u})^{a+n+1}}.\nonumber
		\end{align}
		\begin{IEEEproof}
			The above expression follows by substituting  (\ref{eq_lem_pv_n_pmf}) and (\ref{eq_lem_cond_Pa}) into (\ref{def_eq_Pa}).
		\end{IEEEproof}    
	\end{theorem}
	
	\section{Coverage Probability}
	It can be observed from the previous discussion that after removing the unavailable UAVs form the original point process $\Phi_{\rm u}$, the available UAVs  form a new PPP $\Phi_{\rm u^{'}}$ with density $\lambda_{u}^{'}=P_{\rm a}\lambda_{\rm u}$. Recalling the association policy in Sec. \ref{user_association}, the typical user is served by the UAV in its hotspot center if it is available, otherwise, it associates with the nearest LoS/NLoS UAV or active charging station, whichever provides the strongest average received power. 
	As stated earlier, the locations of LoS and NLoS available UAVs are modeled by the PPPs $\Phi_{\rm u^{'}_{l}}$, $\Phi_{\rm u^{'}_{n}}$, the location of the typical UAV is modeled by $\Phi_{\rm u_{o}}$ (which equals to $\emptyset$ when the typical UAV is unavailable), the locations of the active charging stations (excluding the typical charging station) are modeled by the PPP $\Phi_{\rm c,a}$, and the location of the typical charging station is modeled by $C_{\rm Rs}$. 
	
	In order to compute the coverage probability, the distance distribution to the nearest point in each of these point processes is required, as well as the joint distributions between some of them, as will be clarified in the following part of the paper.
	\begin{lemma}[Distance Distribution]\label{Lemma_distance}
		The probability density function of the distances between the typical user and the UAV in its hotspot center, the nearest available NLoS and LoS UAV, denoted by $f_{\rm R_{u_o}}(r)$, $f_{\rm R_{u^{'},n}}(r)$ and $f_{\rm R_{u^{'},l}}(r)$, respectively, are given by
		\begin{align}
		f_{\rm R_{u_o}}(r) &= \frac{2r}{r_{c}^2}, \quad h\leq r \leq \sqrt{r_c^2+h^2},\label{dist_u_o}\\
		f_{\rm R_{u^{'},n}}(r) &= 2\pi\lambda_{\rm u}^{'}P_{\rm n}(r)r\exp\bigg(-2\pi\lambda_{\rm u}^{'}\int_{0}^{\sqrt{r^2-h^2}}zP_{\rm n}(\sqrt{z^2+h^2}){\rm d}z\bigg),\label{dist_u_p_n}\\
		f_{\rm R_{u^{'},l}}(r) &= 2\pi\lambda_{\rm u}^{'}P_{\rm l}(r)r\exp\bigg(-2\pi\lambda_{\rm u}^{'}\int_{0}^{\sqrt{r^2-h^2}}zP_{\rm l}(\sqrt{z^2+h^2}){\rm d}z\bigg)\label{dist_u_p_l},
		\end{align}
		where $P_{\rm n}(r)$ and $P_{\rm l}(r)$ are defined in (\ref{pl_pn}). Recall that $R_{\rm c}$ and $R_{\rm s}$ are the distances from the cluster origin to the nearest active charging station $C_{\rm R_c}$ and the typical charging station $C_{\rm R_s}$, respectively. Note that $R_{\rm c}$ is greater than $R_{\rm s}$ by construction. The PDF of $R_{\rm c}$ is 
		\begin{align}
		f_{\rm R_{c}}(r|R_{s}) = \frac{2\pi\lambda_{\rm c}^{'}r\exp(-\pi\lambda_{\rm c}^{'}r^2)}{\exp(-\pi\lambda_{\rm c}^{'}R_{\rm s}^2)},\ r\geq R_{\rm s}.\label{eq_dist_Rc}
		\end{align}
		
		Let $R_{\rm su}$ and $R_{\rm cu}$ be the distances between the typical user and $C_{\rm R_s}$ and  $C_{\rm R_c}$, respectively. Conditioned on $R_{\rm s}$ and $R_{\rm c}$, their PDFs are given by
		\begin{align}
		\label{eq_dist_Rsu}
		f_{\rm R_{\{su,cu\}}}(r|R_{\rm \{s,c\}})&=\left\{ 
		\begin{aligned}
		\frac{2r}{r_c^2},  & \quad \text{\rm if} \quad 0<r<r_c-R_{\rm \{s,c\}},\\
		\frac{2r}{\pi r_c^2}\arccos\bigg(\frac{r^2-r_c^2+R_{\rm \{s,c\}}^2}{2R_{\rm \{s,c\}}r}\bigg),  & \quad\text{\rm if}\quad r_c-R_{\rm \{s,c\}}<r<R_{\rm \{s,c\}}+r_c,\\
		\end{aligned} \right.
		\end{align}
		where $R_{\rm \{s,c\}}<r_c$. Otherwise,
		\begin{align}
		f_{\rm R_{\{su,cu\}}}(r|R_{\rm \{s,c\}})&=
		\frac{2r}{\pi r_c^2}\arccos\bigg(\frac{r^2-r_c^2+R_{\rm \{s,c\}}^2}{2R_{\rm \{s,c\}}r}\bigg),  &\quad\text{\rm if} \quad R_{\rm \{s,c\}}-r_c<r<R_{\rm \{s,c\}}+r_c.\label{eq_dist_Rcu}
		\end{align}
		Hence, the CDFs of $R_{\rm su}$ and $R_{\rm cu}$ are given by
		\begin{small}
			\begin{align}
			\label{eq_cdf_dist_Rsu}
			F_{\rm  R_{\{su,cu\}}}(r|R_{\rm \{s,c\}})&=\left\{ 
			\begin{aligned}
			\frac{r^2}{r_c^2},  & \quad \text{\rm if} \quad 0<r<r_c-R_{\rm \{s,c\}},\\
			\frac{\int_{r_c-R_{\rm \{s,c\}}}^{r}2x\arccos\bigg(\frac{x^2-r_c^2+R_{\rm \{s,c\}}^2}{2R_{\rm \{s,c\}}x}\bigg){\rm d}x}{\pi r_c^2}+\frac{(r_c-R_{\rm \{s,c\}})^2}{r_c^2},& \quad\text{\rm if} \quad r_c-R_{\rm \{s,c\}}<r<R_{\rm \{s,c\}}+r_c,\\
			1, &\quad\text{\rm if} \quad R_{\rm \{s,c\}}+r_c<r,\\
			\end{aligned} \right.
			\end{align}
		\end{small}
		when $R_{\rm \{s,c\}}<r_c$. Otherwise,
		\begin{align}
		\label{eq_cdf_dist_Rcu}
		F_{\rm R_{\{su,cu\}}}(r|R_{\rm \{s,c\}})&=\left\{ 
		\begin{aligned}
		\frac{\int_{R_{\rm \{s,c\}}-r_c}^{r}2x\arccos\bigg(\frac{x^2-r_c^2+R_{\rm \{s,c\}}^2}{2R_{\rm \{s,c\}}x}\bigg){\rm d}x}{\pi r_c^2},  &\quad\text{\rm if} \quad R_{\rm \{s,c\}}-r_c<r<R_{\rm \{s,c\}}+r_c,\\
		1, &\quad\text{\rm if} \quad R_{\rm \{s,c\}}+r_c<r.
		\end{aligned} \right.
		\end{align}
	\end{lemma}
Now that we have derived all the required distance distributions, in the following part, we aim to characterize the association probability with each of the UAVs and the active charging stations when the the typical UAV is unavailable. 	
	
	
	\begin{lemma}[Associate Probability]\label{lem_associate}
		Let $\mathcal{A}_{\rm LoS}(r)$, $\mathcal{A}_{\rm NLoS}(r)$, $\mathcal{A}_{\rm Cs}(r)$ and $\mathcal{A}_{\rm Cc}(r)$ be the probabilities that the typical user associates with the nearest LoS, NLoS UAV, $C_{\rm R_s}$ and $C_{\rm R_c}$ at distance $r$,  respectively. When $C_{\rm R_s}$ is active, the association probabilities are given by
		\begin{align}
		\mathcal{A}_{\rm LoS,a}(r|R_{\rm s},R_{\rm c})&=\mathcal{A}_{\rm LoS-NLoS}(r)\mathcal{A}_{\rm LoS-Cs}(r|R_{\rm s})\mathcal{A}_{\rm LoS-Cc}(r|R_{\rm c}),\nonumber\\
		\mathcal{A}_{\rm NLoS,a}(r|R_{\rm s},R_{\rm c})&=\mathcal{A}_{\rm NLoS-LoS}(r)\mathcal{A}_{\rm NLoS-Cs}(r|R_{\rm s})\mathcal{A}_{\rm NLoS-Cc}(r|R_{\rm c}),\nonumber\\
		\mathcal{A}_{\rm Cs}(r|R_{\rm s},R_{\rm c})&=\mathcal{A}_{\rm Cs-LoS}(r|R_{\rm s})\mathcal{A}_{\rm C_s-NLoS}(r|R_{\rm s})\mathcal{A}_{\rm Cs-Cc}(r|R_{\rm s},R_{\rm c}),\nonumber\\
		\mathcal{A}_{\rm Cc,a}(r|R_{\rm s},R_{\rm c})&=\mathcal{A}_{\rm Cc-LoS}(r|R_{\rm c})\mathcal{A}_{\rm Cc-NLoS}(r|R_{\rm c})\mathcal{A}_{\rm Cc-Cs}(r|R_{\rm s},R_{\rm c}).\nonumber
		\end{align}
		When $C_{\rm R_s}$ is not active, the association probabilities are given by
		\begin{align}
		\mathcal{A}_{\rm LoS,n}(r|R_{\rm c})&=\mathcal{A}_{\rm LoS-NLoS}(r)\mathcal{A}_{\rm LoS-Cc}(r|R_{\rm c}),\nonumber\\
		\mathcal{A}_{\rm NLoS,n}(r|R_{\rm c})&=\mathcal{A}_{\rm NLoS-LoS}(r)\mathcal{A}_{\rm NLoS-Cc}(r|R_{\rm c}),\nonumber\\
		\mathcal{A}_{\rm Cc,n}(r|R_{\rm c})&=\mathcal{A}_{\rm Cc-LoS}(r|R_{\rm c})\mathcal{A}_{\rm Cc-NLoS}(r|R_{\rm c}),\nonumber
		\end{align}
		in which,
		\begin{align}
		\mathcal{A}_{\rm LoS-NLoS}(r)
		&= \exp\bigg(-2\pi\lambda_{\rm u}^{'}\int_{0}^{\sqrt{d_{\rm n}^{2}(r)-h^2}}zP_{\rm n}(\sqrt{z^2+h^2}){\rm d}z\bigg),\nonumber\\
		\mathcal{A}_{\rm NLoS-LoS}(r) &= \exp\bigg(-2\pi\lambda_{\rm u}^{'}\int_{0}^{\sqrt{d_{\rm l}^{2}(r)-h^2}}zP_{\rm l}(\sqrt{z^2+h^2}){\rm d}z\bigg),\nonumber\\
		\mathcal{A}_{\rm \{LoS,NLoS\}-\{Cs,Cc\}}(r|R_{\rm \{s,c\}}) &= 1-F_{\rm R_{\{su,cu\}}}(D_{\rm \{l,n\}}(r|R_{\rm \{s,c\}})),\nonumber\\
		\mathcal{A}_{\rm \{Cs,Cc\}-\{Cc,Cs\}}(r|R_{\rm \{c,s\}}) &= 1-F_{\rm R_{\{cu,su\}}}(r|R_{\rm \{c,s\}}),\nonumber\\
\mathcal{A}_{\rm \{Cs,Cc\}-\{LoS,NLoS\}}(r|R_{\rm \{s,c\}}) &= \exp\bigg(-2\pi\lambda_{\rm u}^{'}\int_{0}^{\sqrt{\hat{D}_{\rm \{l,n\}}^2(r)-h^2}}zP_{\rm \rm \{l,n\}}(\sqrt{z^2+h^2}){\rm d}z\bigg),\nonumber
		\end{align}
		where $D_{\rm \{l,n\}}(r) = r^{\frac{\alpha_{\rm \{l,n\}}}{\alpha_{\rm t}}}\eta_{\rm \{l,n\}}^{\frac{1}{\alpha_t}}$, $\hat{D}_{\rm l}(r) = \max\left(h,r^{\frac{\alpha_{\rm t}}{\alpha_{\rm l}}}\eta_{\rm l}^{\frac{1}{\alpha_{\rm l}}}\right)$, $\hat{D}_{\rm n}(r) =\max\left(h,r^{\frac{\alpha_{\rm t}}{\alpha_{\rm n}}}\eta_{\rm n}^{\frac{1}{\alpha_{\rm n}}}\right)$,
			\begin{align}
	\label{eq_d_L}
	d_{\rm l}(r) = \bigg(\frac{\eta_{\rm l}}{\eta_{\rm n}}\bigg)^{\frac{1}{\alpha_{\rm l}}}r^{\frac{\alpha_{\rm n}}{\alpha_{\rm l}}},
	\end{align}
	and
	\begin{align}
	\label{eq_d_N}
	d_{\rm n}(r) = \max\bigg(h,\bigg(\frac{\eta_{\rm n}}{\eta_{\rm l}}\bigg)^{\frac{1}{\alpha_{\rm n}}}r^{\frac{\alpha_{\rm l}}{\alpha_{\rm n}}}\bigg).
	\end{align}
		\begin{IEEEproof}
			See Appendix \ref{app_associate}.
		\end{IEEEproof}
	\end{lemma}
	The final requirement to derive the coverage probability is the Laplace transform of the aggregate interference, which  is provided in the following lemma.
	\begin{lemma}[Laplace Transform of Interference]\label{lem_Laplace}
		Using subscripts $a$ and $n$ to capture the events of the typical charging station is active and inactive, respectively, the Laplace transform of the interference power conditioned on the serving UAV $u_s$ located at $x$ is
		\begin{align}
		\mathcal{L}_{\rm I,\{a,n\}}(s,\|x\|) &=\exp\biggl(-2\pi\lambda_{\rm u}^{'}\int_{a(\|x\|)}^{\infty}\bigg[1-\bigg(\frac{m_{\rm n}}{m_{\rm n}+s\eta_{\rm n}\rho_{\rm u}(z^2+h^2)^{-\frac{\alpha_{\rm n}}{2}}}\bigg)^{m_{\rm n}}\bigg]zP_{\rm n}(\sqrt{z^2+h^2}){\rm d}z\biggl)\nonumber\\
		& \times \exp\biggl(-2\pi\lambda_{\rm u}^{'}\int_{b(\|x\|)}^{\infty}\bigg[1-\bigg(\frac{m_{\rm l}}{m_{\rm l}+s\eta_{\rm l}\rho_{\rm u}(z^2+h^2)^{-\frac{\alpha_{\rm l}}{2}}}\bigg)^{m_{\rm l}}\bigg]zP_{\rm l}(\sqrt{z^2+h^2}){\rm d}z\biggl)\nonumber\\
		&\times \exp\biggl(-2\pi\lambda_{\rm c}^{'}\int_{c(\|x\|)}^{\infty}\bigg[1-\bigg(\frac{1}{1+s\rho_{\rm u}z^{-\alpha_{\rm t}}}\bigg)\bigg]z{\rm d}z\biggl),\nonumber
		\end{align}
		in which,
		\begin{align}
		a(\|x\|)=\left\{ 
		\begin{aligned}
		0,  & \quad \text{\rm if} \quad u_s \in \Phi_{\rm u_{o}},\\
		\sqrt{d_{\rm n}^2(\|x\|)-h^2},  & \quad \text{\rm if} \quad u_s \in \Phi_{\rm u^{'}_{l}},\\
		\sqrt{\|x\|^2-h^2},  & \quad \text{\rm if} \quad u_s \in \Phi_{\rm u_{n}^{'}},\\
		\sqrt{\hat{D}_{ n}^2(\|x\|)-h^2},  & \quad \text{\rm if} \quad u_s \in \Phi_{\rm c,a}\cup\{C_{\rm R_s}\},\\
		\end{aligned} \right.\nonumber
		\end{align}
		\begin{align}
		b(\|x\|)=\left\{ 
		\begin{aligned}
		0,  & \quad \text{\rm if} \quad u_s \in \Phi_{\rm u_{o}},\\
		\sqrt{\|x\|^2-h^2},  & \quad \text{\rm if} \quad u_s \in \Phi_{\rm u^{'}_{l}},\\
		\sqrt{d_{\rm l}^2(\|x\|)-h^2},  & \quad \text{\rm if} \quad u_s \in \Phi_{\rm u_{n}^{'}},\\
		\sqrt{\hat{D}_{ l}^2(\|x\|)-h^2},  & \quad \text{\rm if} \quad u_s \in \Phi_{\rm c,a}\cup\{C_{\rm R_s}\},\\
		\end{aligned} \right.\nonumber
		\end{align}
				\begin{align}
		c(\|x\|)=\left\{ 
		\begin{aligned}
		0,  & \quad \text{\rm if} \quad u_s \in \Phi_{\rm u_{o}},\\
		{D}_{ l}(\|x\|),  & \quad \text{\rm if} \quad u_s \in \Phi_{\rm u^{'}_{l}},\\
		{D}_{ n}(\|x\|),  & \quad \text{\rm if} \quad u_s \in \Phi_{\rm u_{n}^{'}},\\
		R_{\rm cu},  & \quad \text{\rm if} \quad u_s \in \Phi_{\rm c,a}\cup\{C_{\rm R_s}\} \quad \text{\rm and} \quad C_{\rm R_s}\quad \text{\rm is inactive},\\
		\min(R_{\rm su},R_{\rm cu}),  & \quad \text{\rm if} \quad u_s \in \Phi_{\rm c,a}\cup\{C_{\rm R_s}\} \quad \text{\rm and} \quad C_{\rm R_s}\quad \text{\rm is active},\\
		\end{aligned} \right.\nonumber
		\end{align}
		\begin{IEEEproof}
			See Appendix \ref{app_Laplace}.
		\end{IEEEproof}
	\end{lemma}
	
	Now that we have developed expressions for the relevant distances and the association probabilities, we study the coverage probability as explained in Definition \ref{def_cov}.
	\begin{theorem}[Coverage Probability]\label{lem_P_cov}
	The coverage probability, provided in (\ref{cov:eq}), can be computed using the following expressions for the coverage probability when the typical UAV is available and unavailable, respectively:
		\begin{align}
		P_{\rm cov,U_o} &= P_{\rm cov,U_{o,l}}+P_{\rm cov,U_{o,n}},\nonumber\\
		P_{\rm cov,\bar{U}_o}&=P_{\rm cov,U^{'}_{l}}+P_{\rm cov,U^{'}_{n}}+P_{\rm cov,Cs}+P_{\rm cov,Cc},\nonumber
		\end{align}
		where,
		\begin{align}
		P_{\rm cov,U_{o,l}} =& \mathbb{E}_{\rm \{R_s,R_c,R_{cu},R_{su}\}}\bigg[\int_{h}^{\sqrt{h^2+r_{c}^{2}}}\sum_{k=0}^{m_{\rm l}-1}\frac{(-m_{\rm l}g_{\rm l}(r))^k}{k!}\nonumber\\
		&\times\bigg[P_{\rm C,a}\frac{\partial^{k}}{\partial s^{k}}\mathcal{L}_{\rm \sigma^2+I,a}(s,r)+(1-P_{\rm C,a})\frac{\partial^{k}}{\partial s^{k}}\mathcal{L}_{\rm\sigma^2+I,n}(s,r)\bigg]_{s=m_{\rm l}g_{\rm l}(r)}P_{\rm l}(r) \frac{2r}{r_{c}^2}{\rm d}r\bigg],\label{cov_U_o_L}\\
		P_{\rm cov,U_{o,n}} =& \mathbb{E}_{\rm \{R_s,R_c,R_{cu},R_{su}\}}\bigg[\int_{h}^{\sqrt{h^2+r_{c}^{2}}}\sum_{k=0}^{m_{\rm n}-1}\frac{(-m_{\rm n}g_{\rm n}(r))^k}{k!}\nonumber\\
		&\times\bigg[P_{\rm C,a}\frac{\partial^{k}}{\partial s^{k}}\mathcal{L}_{\rm \sigma^2+I,a}(s,r)+(1-P_{\rm C,a})\frac{\partial^{k}}{\partial s^{k}}\mathcal{L}_{\rm\sigma^2+I,n}(s,r)\bigg]_{s=m_{\rm n}g_{\rm n}(r)}P_{\rm n}(r) \frac{2r}{r_{c}^2}{\rm d}r\bigg],\\
		P_{\rm cov,U^{'}_{l}} =& \mathbb{E}_{\rm \{R_s,R_c,R_{cu},R_{su}\}}\bigg[\int_{h}^{\infty}\sum_{k=0}^{m_{\rm l}-1}\frac{(-m_{\rm l}g_{\rm l}(r))^k}{k!}\bigg(\mathcal{A}_{\rm LoS,a}(r|R_{\rm s},R_{\rm c})P_{\rm Crs,a}\frac{\partial^{k}}{\partial s^{k}}\mathcal{L}_{\rm \sigma^2+I,a}(s,r) \nonumber\\
		&+ \mathcal{A}_{\rm LoS,n}(r|R_{\rm c})(1-P_{\rm Crs,a}) \frac{\partial^{k}}{\partial s^{k}}\mathcal{L}_{\rm\sigma^2+I,n}(s,r)\bigg)_{s=m_{\rm l}g_{\rm l}(r)}f_{\rm R_{u^{'},l}}(r) {\rm d}r\bigg],\label{cov_U_p_L}\\
		P_{\rm cov,U^{'}_{n}} =& \mathbb{E}_{\rm \{R_s,R_c,R_{cu},R_{su}\}}\bigg[\int_{h}^{\infty}\sum_{k=0}^{m_{\rm n}-1}\frac{(-m_{\rm n}g_{\rm n}(r))^k}{k!}\bigg(\mathcal{A}_{\rm NLoS,a}(r|R_{\rm s},R_{\rm c})P_{\rm Crs,a} \frac{\partial^{k}}{\partial s^{k}}\mathcal{L}_{\rm\sigma^2+I,a}(s,r) \nonumber\\
		&+ \mathcal{A}_{\rm NLoS,n}(r|R_{\rm c})(1-P_{\rm Crs,a}) \frac{\partial^{k}}{\partial s^{k}}\mathcal{L}_{\rm \sigma^2+I,n}(s,r)\bigg)_{s=m_{\rm n}g_{\rm n}(r)}f_{\rm R_{u^{'},n}}(r){\rm d}r\bigg],\\
		P_{\rm cov,Cs} =& \mathbb{E}_{\rm \{R_s,R_c,R_{cu}\}}\bigg[\int_{0}^{\infty}\bigg(\mathcal{A}_{\rm Cs}({r}|R_{\rm s},R_{\rm c})P_{\rm Crs,a} \mathcal{L}_{\rm\sigma^2+I,a}(\theta\rho_{\rm u}^{-1} {r}^{\alpha_{\rm T}},{r})\bigg)f_{\rm R_{su}}(r|R_{s}){\rm d}r\bigg],\\
		P_{\rm cov,Cc} =& \mathbb{E}_{\rm \{R_s,R_c,R_{su}\}}\bigg[\int_{0}^{\infty}\bigg(\mathcal{A}_{\rm Cc,a}(r|R_{\rm s},R_{\rm c})P_{\rm Crs,a} \mathcal{L}_{\rm\sigma^2+I,a}(\theta\rho_{\rm u}^{-1} r^{\alpha_{\rm T}},r) \nonumber\\
		&+\mathcal{A}_{\rm Cc,n}(r|R_{\rm c})(1-P_{\rm Crs,a}) \mathcal{L}_{\rm\sigma^2+I,n}(\theta\rho_{\rm u}^{-1} r^{\alpha_{\rm T}},r)\bigg)f_{\rm R_{cu}}(r|R_{c}){\rm d}r\bigg],
		\end{align}
		in which
		\begin{align}
		g_{\rm l}(r) &= \theta\eta_{\rm l}^{-1}\rho_{\rm u}^{-1}r^{\alpha_{\rm l}}, \nonumber\\
		g_{\rm n}(r) &= \theta\eta_{\rm n}^{-1}\rho_{\rm u}^{-1}r^{\alpha_{\rm n}} ,
		\label{eq_g_l_g_n}
		\end{align}
$		f_{\rm R_{c}}(r|R_{s})$ is given in (\ref{eq_dist_Rc}), and $f_{\rm R_{s}}(r)=2\lambda_c\pi r\exp(-\lambda_c\pi r^2)$.
		\begin{IEEEproof}
			See Appendix \ref{app_P_cov}.
		\end{IEEEproof}
	\end{theorem}
	Note that the summations and derivatives in Theorem \ref{lem_P_cov} are obtained from
	$$\frac{\Gamma_{u}(m,mg)}{\Gamma(m)}=\exp(-mg)\sum_{k=0}^{m-1}\frac{(mg)^k}{k!},$$ and $$\mathbb{E}_{U}[\exp(-sU)U^{k}]=(-1)^k\frac{\partial^{k}}{\partial s^{k}}\mathcal{L}_{U}(s).$$
	
	Observe that the above expressions require evaluating higher order derivatives of the Laplace transform. Now we present an approximation for the coverage probability using the upper bound of the CDF of the Gamma distribution \cite{alzer1997some}.
	\begin{lemma}[Approximated Coverage Probability] \label{lem_approximation}
		$P_{\rm cov,U_{o,\{l,n\}}}$ and $P_{\rm cov,U^{'}_{\{l,n\}}}$ can be approximated by using the upper bound of the CDF of the Gamma distribution as
		\begin{align}
		&P_{\rm cov,U_{o,\{l,n\}}} = \mathbb{E}_{\rm \{R_s,R_c,R_{cu},R_{su}\}}\bigg[\sum_{k=1}^{m_{\rm \{l,n\}}}\binom{m_{\rm \{l,n\}}}{k}(-1)^{k+1}\int_{h}^{\sqrt{h^2+r_{c}^2}}P_{\rm \{l,n\}}(r)\frac{2r}{r_c^2}\times\nonumber\\&\bigg(P_{\rm C,a}\mathcal{L}_{\rm \sigma^2+I,a}(k\beta_{2}m_{\rm \{l,n\}}g_{\rm \{l,n\}}(r),r)
		+(1-P_{\rm C,a})\mathcal{L}_{\rm \sigma^2+I,n}(k\beta_{2}m_{\rm \{l,n\}}g_{\rm \{l,n\}}(r),r)\bigg){\rm d}r\bigg],\\
		&P_{\rm cov,U^{'}_{\{l,n\}}} =  \mathbb{E}_{\rm \{R_s,R_c,R_{cu},R_{su}\}}\bigg[\sum_{k=1}^{m_{\rm {\{l,n\}}}}\binom{m_{\rm {\{l,n\}}}}{k}(-1)^{k+1}\int_{h}^{\infty}f_{\rm R_{u^{'},{\{l,n\}}}}(r)\bigg(\mathcal{A}_{\rm {\{L,NL\}}oS,a}(r|R_{\rm s},R_{\rm c})\times\nonumber\\
		&P_{\rm Crs,a}\mathcal{L}_{\rm \sigma^2+I,a}(k\beta_{2}m_{\rm l}g_{\rm {\{l,n\}}}(r),r)+\mathcal{A}_{\rm {\{L,NL\}}oS,n}(r|R_{\rm c})(1-P_{\rm Crs,a})\mathcal{L}_{\rm \sigma^2+I,n}(k\beta_{2}m_{\rm l}g_{\rm {\{l,n\}}}(r),r)\bigg) {\rm d}r\bigg],
		\end{align}
		in which $\beta_{2}=(m_{\rm {\{l,n\}}}!)^{-\frac{1}{m_{\rm {\{l,n\}}}}}$.
		\begin{IEEEproof}
			See Appendix \ref{app_approximation}.
		\end{IEEEproof}
	\end{lemma}
	\begin{remark}
	The results in Lemma \ref{lem_approximation} efficiently reduce the complexity of computing the coverage probability, since it only requires a simple integral and finite summations. In addition, note that the fading gain of NLoS is $1$ in our system model, we only need to evaluate the integral and no summations. Besides, it is a tight approximation, whose gap can be ignored, more details will be shown in Section \ref{numerical result}.
\end{remark}

	\section{Numerical Result}\label{numerical result}
	
	In this section, we validate our analytical results with simulations and evaluate the impact of various system parameters on the network performance such as the $Ratio=\frac{\lambda_{\rm u}}{\lambda_{\rm c}}$ and the charging station capacity $c$. 
	Unless stated otherwise, we use the simulation parameters as listed herein Table \ref{par_val}. 
	
	\begin{table*}[ht]\caption{Table of Parameters}\label{par_val}
		\centering
		\begin{center}
			\resizebox{\textwidth}{!}{
				\renewcommand{\arraystretch}{1}
				\begin{tabular}{ {c} | {c} | {c}  }
					\hline
					\hline
					\textbf{Parameter} & \textbf{Symbol} & \textbf{Simulation Value}  \\ \hline
					Charging station density & $\lambda_{\rm c}$ & $5\times 10^{-7}$ m$^{-2}$ \\ \hline
					UAV to charging station density ratio & $Ratio$ & 1 to 20 \\ \hline
					Charging station capacity & $c$ & 1 to 6\\ \hline
					PV cell fitting parameters & $a,b$ & 3.5\\ \hline
					Landing/taking off energy & $E_l$ & 2184 J \\ \hline
					Traveling-related power & $P_{\rm m}$ & 161.8 W \\\hline
					Service-related power & $P_{\rm s}$ & 177.5 W \\\hline
					UAV altitude & $h$ & 60 m\\\hline
					UAV velocity & $V$ & 18.46 m/s \\\hline
					Acceleration & $a_{\rm ave}$ & 3.24 m/s$^2$ \\\hline
					Battery capacity & $B_{\rm max}$ & 88.8 W$\cdot$H \\\hline
					Charging time & $T_{\rm ch}$ & 5 min\\\hline
					Radius of MCP disk & $r_c$ & 120 m \\\hline
					N/LoS environment variable & $A, B$ & 25.27,0.5 \\\hline
					Transmission power & $\rho_{\rm u}$ & 0.2 W\\\hline
					SINR threshold & $\beta$ & 0 dB \\\hline
					Noise power & $\sigma^2 $ & $10^{-9}$ W\\\hline
					N/LoS and active charging station path-loss exponent & $\alpha_{\rm n},\alpha_{\rm l},\alpha_{\rm t}$ & $4,2.1,4$ \\\hline
					N/LoS fading gain & $m_{\rm n},m_{\rm l}$ & $1,3$ \\\hline
				    N/LoS additional loss& $\eta_{\rm n},\eta_{\rm l}$ & $20,0$ dB 
				    \\\hline\hline
			\end{tabular}}
		\end{center}
	\end{table*}
	
	\begin{figure}[ht]
		\centering
		\includegraphics[width=0.7\columnwidth]{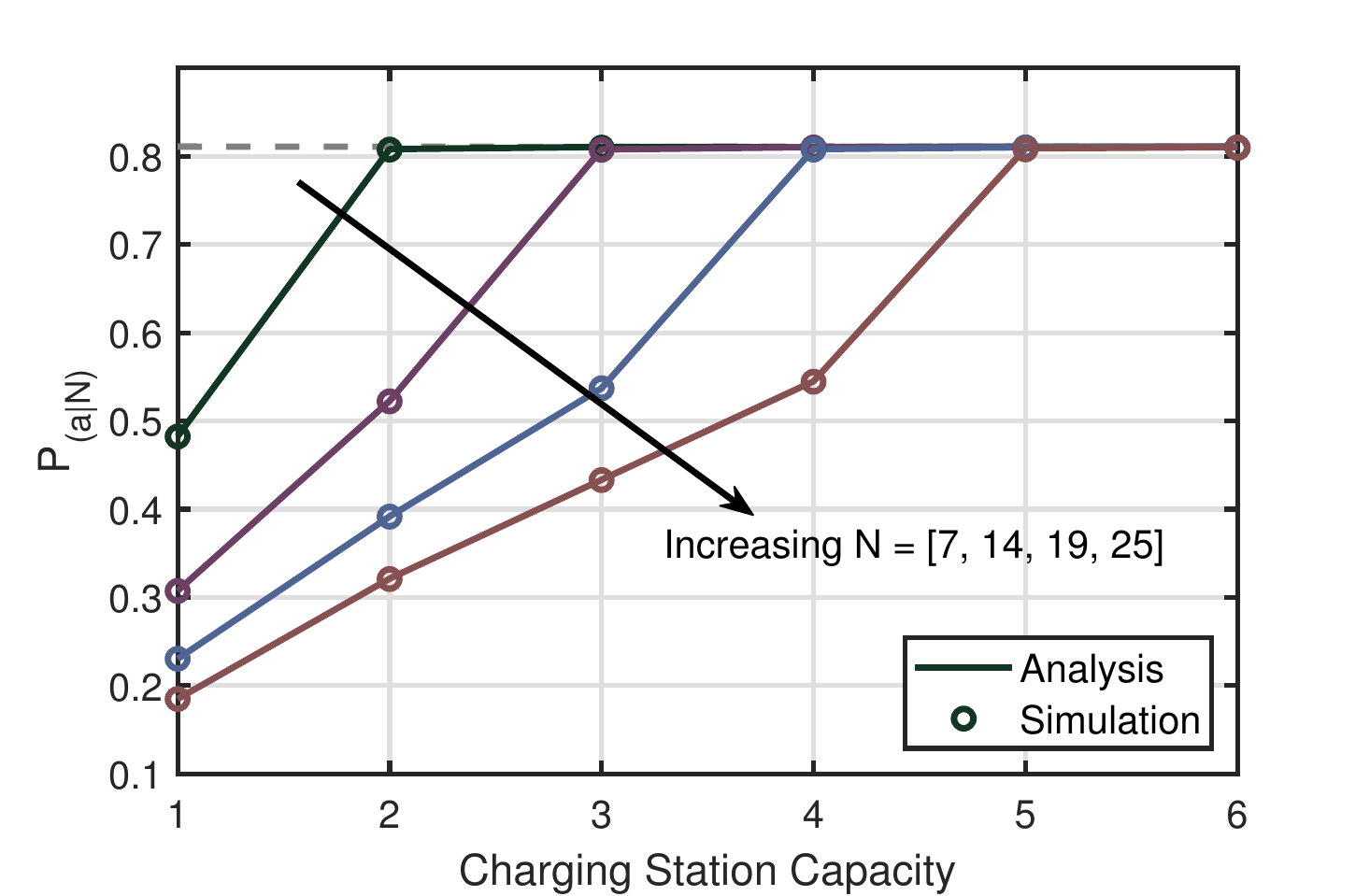}
		\caption{Conditioned availability probability of UAVs for different values of the charging station capacity for different values of $N$. Dash line denotes the maximum achievable availability probability in which there is no waiting time at the charging station.}
		\label{Fig_pa_n}
	\end{figure}
	
	In Fig.~\ref{Fig_pa_n} we plot the availability probability conditioned on the number of UAVs in the typical PV cell against the charging station capacity $c$. For a given value of $N$, the availability probability increases with the charging station capacity until the maximum achievable value in which waiting time equals to 0 min.  
	
	\begin{figure}[ht]
		\centering
		\includegraphics[width=0.7\columnwidth]{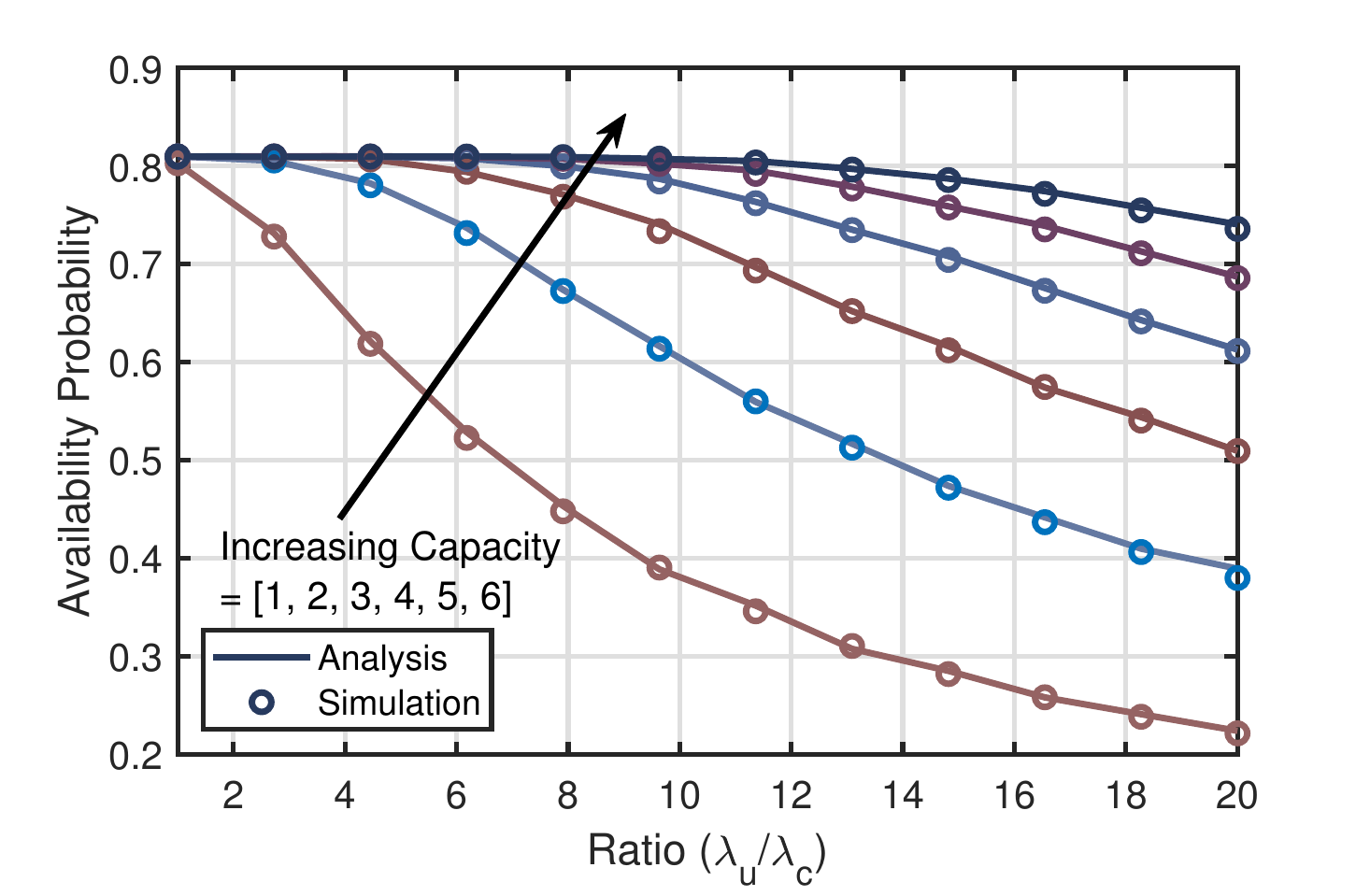}
		\caption{Probability of availability against different values of the $Ratio=\frac{\lambda_u}{\lambda_c}$ for different charging station capacities.}
		\label{Fig_pa} 
	\end{figure}
	
	The results in Fig.~\ref{Fig_pa} reveal that charging station capacity has a significant impact on the availability of UAVs. When the capacity is low, the availability probability of UAVs decreases dramatically with the increase in the $Ratio=\frac{\lambda_{\rm u}}{\lambda_{\rm c}}$. 
	We also notice that at high values of the $\frac{\lambda_{\rm u}}{\lambda_{\rm c}}$, say $20$, a slight increases in the capacity, from $1$ to $2$, leads to doubling the availability probability from $0.2$ to $0.4$.
	
	\begin{figure}[ht]
		\centering
		\includegraphics[width=0.7\columnwidth]{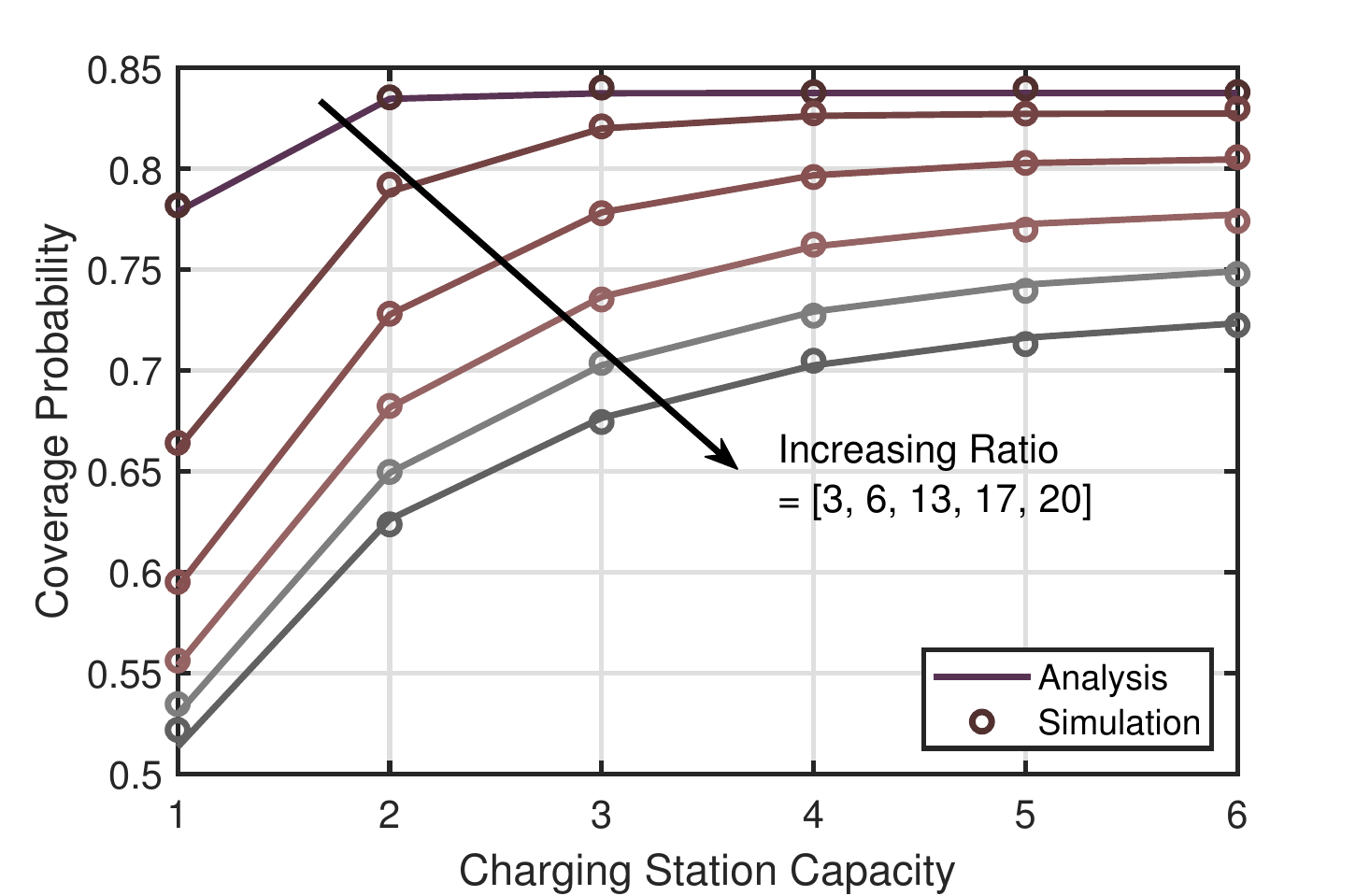}
		\caption{Coverage probability against different values of the charging station capacity for different values of $Ratio=\frac{\lambda_u}{\lambda_c}$.}
		\label{Fig_p_cov_n}
	\end{figure}
	
\begin{figure}[ht]
		\centering
		\includegraphics[width=0.7\columnwidth]{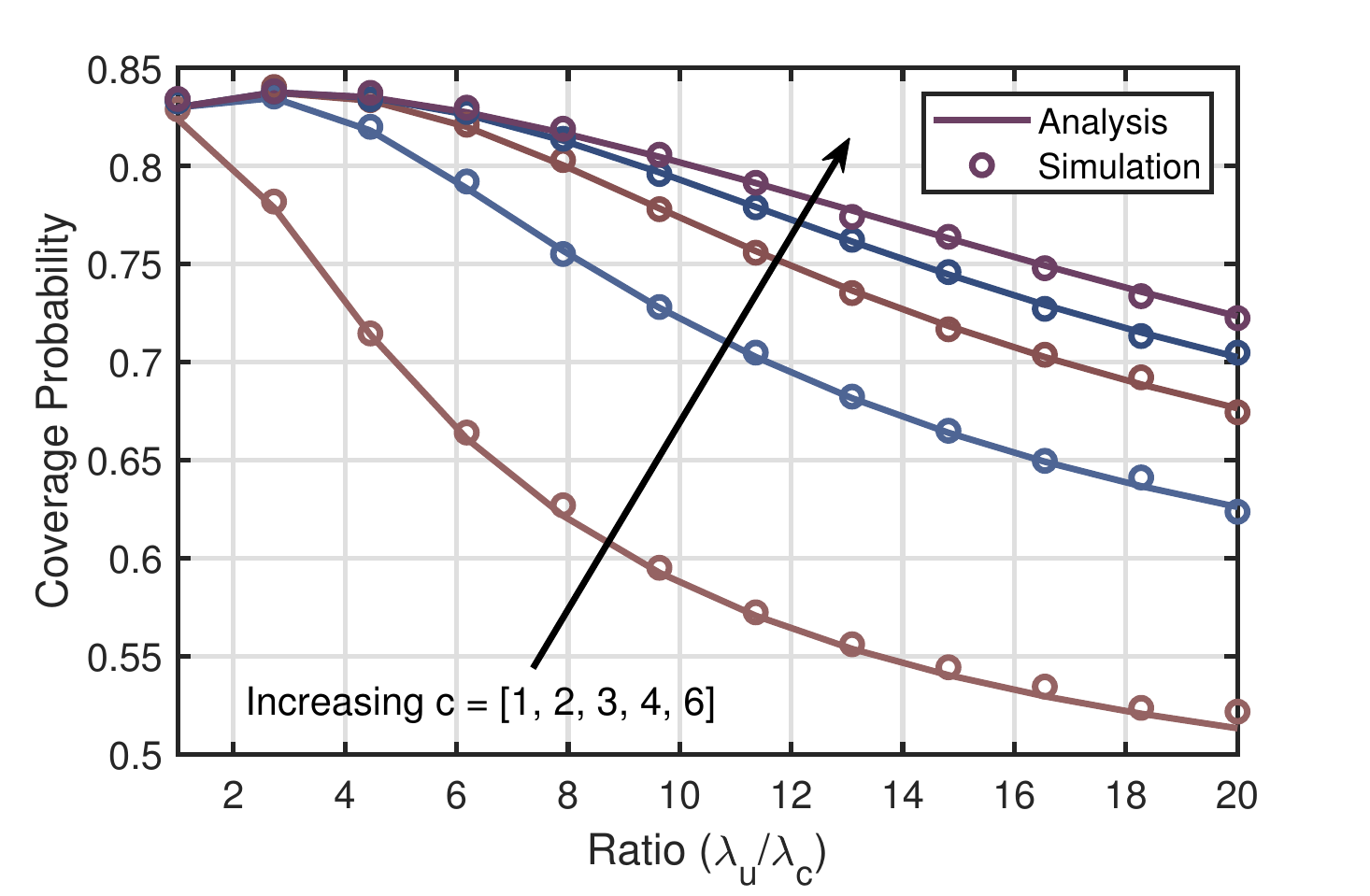}
		\caption{Coverage probability against different values of $Ratio=\frac{\lambda_u}{\lambda_c}$ for different charging station capacities.}
		\label{Fig_p_cov}
	\end{figure}

	As can be seen in Fig.~\ref{Fig_p_cov_n}, the charging station capacity and the $Ratio=\frac{\lambda_u}{\lambda_c}$ have a huge impact on the coverage probability. In addition, we observe that the coverage probability becomes less sensitive to the changes of charging station capacity beyond a certain point, due to achieving the zero waiting time. The same observations also applies in Fig.~\ref{Fig_p_cov}.
	
Fig.~\ref{Fig_p_cov} shows the influence of the $Ratio=\frac{\lambda_u}{\lambda_c}$ on the coverage probability of the considered setup. We observe 	a sharp decrease in the values of the coverage probability with this ratio at smaller values of the charging station capacity. For instance, at $c=1$, the coverage probability drops from $0.83$ to $0.52$ as we increase the ratio from $1$ to $20$, which is consistent with sharp decrease in availability probability. However, the influence becomes much less as we increase the charging station capacity.       Coverage probability increases first due to the fact that the typical user is more likely to establish a LoS link with serving UAV. However, if we continue increasing the density of UAVs, availability probability decreases quickly owing to a long queue and waiting time. Besides  higher interference, users need to connect to a nearby UAV or an active charging station, which provides a worse signal compared with the one in its hotspot UAV.

These results reveal an interesting trade-off between deploying high density of charging stations with small capacity or deploying low density of charging stations with large capacity.

\section{Conclusion}
In this paper, we have studied a novel setup that captures the influence of the limited UAV's battery capacity on the performance of the a UAV-enabled wireless network. Firstly, we derived the availability probability of a UAV as a function of the battery size, the charging time, the density of UAVs, the density of the charging stations, and the charging station's capacity. Next, we used the availability probability to compute the overall coverage probability of the considered setup. We have shown how the performance of the considered setup degrades as the capacity of the charging stations decreases or as the ratio between the density of the UAVs and the density of the charging stations increases.

This work tapped a new aspect of the performance of UAV-enabled wireless networks, which can be expanded in various directions. For instance, the performance could be enhanced if each charging station schedules the arrivals and departures of the UAVs to avoid conflicts. The optimal scheduling given the locations of the locations of the UAVs, the charging time, and the power consumption model, is a very interesting open problem. 
	
	\appendix
	\subsection{Proof of Lemma~\ref{lem_pv_n_pmf}}\label{app_pv_n_pmf}
	We adopt the two-parameter gamma function in \cite{FERENC2007518} to fit the area of PV cells 
	\begin{align}
	f(y)=\frac{b^a}{\Gamma(a)}y^{(a-1)}e^{-by},
	\end{align}
	in which, $a$ and $b$ are two fitting parameters. As for a typical UAV, the probability that it locates in the typcial PV cell is proportional to the area of that cell \cite{6497002}. The PDF of the biased area $A_{c}^{'}$ is given by
	\begin{align}
	f_{\rm A_{c}^{'}}(c)&=\frac{c f_{\rm A_{c}}(c)}{\mathbb{E}[A_{c}]}=\frac{b^a}{\Gamma(a)}\lambda_{\rm c}(c\lambda_{\rm c})^{a}e^{-bc\lambda_{\rm c}},
	\end{align}
	in which $	f_{\rm A_c}(c)=\frac{b^a}{\Gamma(a)}\lambda_{\rm c}(c\lambda_{\rm c})^{(a-1)}e^{-bc\lambda_{\rm c}}$ is the PDF of the unbiased area.
	
	The number of UAVs per PV cell is a Poisson distributed random variable, that is
	\begin{align}
	\mathbb{P}(N=n)&=\mathbb{E}_{\rm A_c^{'}}[\mathbb{P}(N=n|A_{c}^{'})] =\int_{0}^{\infty}\mathbb{P}(N=n)f_{\rm A_{c}^{'}}(c){\rm d}c \nonumber\\
	&=\int_{0}^{\infty}\frac{(\lambda_{\rm u} c)^{n}e^{-\lambda_{\rm u} c}}{n!}\frac{b^a}{\Gamma(a)}\lambda_{\rm c}(c\lambda_{\rm c})^{a}e^{-bc\lambda_{\rm c}}{\rm d}c  =\frac{\Gamma(a+n+1)}{\Gamma(a)}\frac{b^a}{n!}\frac{\lambda_{\rm c}^{a+1}\lambda_{\rm u}^{n}}{(b\lambda_{\rm c}+\lambda_{\rm u})^{a+n+1}}.
	\end{align}
	
%
	
	\subsection{Proof of Lemma~\ref{lem_cond_Pa}}\label{app_cond_Pa}
	Let 
	\begin{align}
	\label{eq_Pa_ns}
	P_{\rm (a|N,S)}&=\mathbb{E}_{\rm \Phi_{c}}\bigg[\frac{T_{\rm se}}{T_{\rm se}+T_{\rm ch}+T_{\rm w}(i)+2T_{\rm tra}(x)+2T_{\rm land}}\bigg],\\
	\label{pro_eq_P_NSRs}
	P_{\rm (a|N,S,R_{\rm s})}&=\frac{T_{\rm se}}{T_{\rm se}+T_{\rm ch}+T_{\rm w}(i)+2T_{\rm tra}(x)+2T_{\rm land}},
	\end{align}
	we refer to event $S$ as  conditioned  state of waiting time.
	Substituting (\ref{eq_Tland}), (\ref{eq_Tw}), (\ref{eq_Ttra}) and (\ref{eq_Tse}) into (\ref{pro_eq_P_NSRs}), that is 
	\begin{align}
	P_{\rm (a|N,S,R_{\rm s})}&=\frac{V(B_{\rm max}-2E_l)-2P_{\rm m}R_{\rm s}(x)}{V\bigg(B_{\rm max}-2E_l+P_{\rm s}T_{\rm ch}(1+i)+4P_{\rm s}\sqrt{\frac{2h}{a}}\bigg)+2R_{\rm s}(x)(P_{\rm s}-P_{\rm m})}\nonumber\\
	&=\frac{a_1-a_2R_{\rm s}(x)}{a_3(i)+a_4R_{\rm s}(x)},
	\end{align}
	in which $a_1,a_2,a_3(i)$ and $a_4$ has been defined in Lemma \ref{lem_cond_Pa}.

	The CDF of $P_{\rm (a|N,S,R_{\rm s})}$ is defined as
	\begin{align}
	F_{P_{\rm (a|N,S,R_{\rm s})}}(y)&=\mathbb{P}\bigg(\frac{a_1-a_2R_{\rm s}(x)}{a_3(i)+a_4R_{\rm s}(x)} \leq y\bigg),
	\end{align}
	given that $P_{\rm (a|N,S,R_{\rm s})}$ is a decreasing function of $R_{\rm s}$, the preimage can be obtained as follows
	\begin{align}\label{eq_Pa_rs}
	F_{P_{\rm (a|N,S,R_{\rm s})}}(y)&=\mathbb{P}\bigg(R_{\rm s}(x)\geq\frac{-a_1+a_3(i)y}{-a_2-a_4y}\bigg)\nonumber\\
	&\stackrel{(a)}{=}\int_{\frac{-a_1+a_3(i)y}{-a_2-a_4y}}^{\infty}2\lambda_{\rm c}\pi r e^{-\lambda_{\rm c}\pi r^2}{\rm d}r=\exp\bigg(-\lambda_{\rm c}\pi \bigg(\frac{-a_1+a_3(i)y}{-a_2-a_4y}\bigg)^2\bigg),
	\end{align}
	step $(a)$ follows from the fact that $R_{\rm s}(x)$ is the first contact distance of PPP. Note that due to assuming the typical UAV to be at the origin, $R_{\rm s}(x)=\|x\|$. Substituting (\ref{eq_Pa_rs}) into (\ref{eq_Pa_ns})
	\begin{align}
	P_{\rm (a|N,S)}&=\mathbb{E}_{\rm \Phi_{c}}[P_{\rm (a|N,S,R_{\rm s})}]=\int_{0}^{\infty}1-F_{P_{\rm (a|N,S,R_{\rm s})}}(y){\rm d}y=\int_{a_5}^{\frac{a_1}{a_3(i)}}1-\exp\bigg(-\lambda_{\rm c}\pi \bigg(\frac{-a_1+a_3(i)y}{-a_2-a_4y}\bigg)^2\bigg){\rm d}y.\label{pro_P_aN}
	\end{align}
	Substituting (\ref{pro_P_aN}) and $0\leq i \leq I_{\rm max}-1$ into (\ref{eq_P_aN}) completes the proof.
	
	\subsection{Proof of Lemma~\ref{lem_associate}}\label{app_associate}
	When the typical UAV is unavailable, the typical user associates receives more power from the nearest LoS UAV than the nearest NLos UAV with the following probability:
	\begin{align}
	\mathcal{A}_{\rm LoS-NLoS}(r) &= \mathbb{P}\bigg(\eta_{\rm l}\rho_{\rm u} r^{-\alpha_{\rm l}}> \eta_{\rm n}\rho_{\rm u}R_{\rm u^{'},n}^{-\alpha_{\rm n}} \bigg)= \mathbb{P}\bigg(R_{\rm U^{'},n}>(\frac{\eta_{\rm n}}{\eta_{\rm l}})^{\frac{1}{\alpha_{\rm n}}}r^{\frac{\alpha_{\rm l}}{\alpha_{\rm n}}}\bigg)\nonumber\\
	&= \mathbb{P}\bigg(R_{\rm U^{'},n}>(\frac{\eta_{\rm n}}{\eta_{\rm l}})^{\frac{1}{\alpha_{\rm n}}}r^{\frac{\alpha_{\rm l}}{\alpha_{\rm n}}}\bigg)= \exp\bigg(-2\pi\lambda_{\rm u}^{'}\int_{0}^{\sqrt{d_{\rm n}^{2}(r)-h^2}}zP_{\rm n}(\sqrt{z^2+h^2}){\rm d}z\bigg),
	\end{align}
	in which $d_{\rm n}(r)$ is given in Lemma~\ref{lem_associate}. The proofs of other association probabilities are similar to $\mathcal{A}_{\rm LoS-NLoS}(r)$, therefore omitted here.
	
	\subsection{Proof of Lemma~\ref{lem_Laplace}}\label{app_Laplace}
	The aggregate interference power and its corresponding Laplace transform is  conditioned on the serving UAV $u_s$ located at $x$, given by
	\begin{align}
	&\mathcal{L}_{\rm I,\{a,n\}}(s,\|x\|) = \mathbb{E}_{\rm I}[\exp(-sI)]\nonumber\\
	=& \mathbb{E}_{\rm \Phi_{\rm u^{'}_n}}\bigg[\prod_{N_i\in\Phi_{\rm u^{'}_n}/u_s}\exp(-s\eta_{\rm n}\rho_{\rm u}G_{\rm n}D_{\rm N_i}^{-\alpha_{\rm n}})\bigg]\times\mathbb{E}_{\rm \Phi_{\rm u^{'}_l}}\bigg[\prod_{L_j\in\Phi_{\rm u^{'}_l}/u_s}\exp(-s\eta_{\rm l}\rho_{\rm u}G_{\rm l}D_{\rm L_j}^{-\alpha_{\rm l}})\bigg]\nonumber\\
	&\times\mathbb{E}_{\rm \Phi_{\rm c,a}}\bigg[\prod_{C_k\in\Phi_{\rm c,a}\cup C_{\rm R_s}/u_s}\exp(-s\rho_{\rm u}H D_{\rm C_k}^{-\alpha_{\rm t}})\bigg]\nonumber\\
	=& \mathbb{E}_{\rm \Phi_{\rm u^{'}_n}}\bigg[\prod_{N_i\in\Phi_{\rm u^{'}_n}/u_s}\mathbb{E}_{\rm g_{\rm n}}[\exp(-s\eta_{\rm n}\rho_{\rm u}G_{\rm n}D_{\rm N_i}^{-\alpha_{\rm n}})]\bigg]\times\mathbb{E}_{\rm \Phi_{\rm u^{'}_l}}\bigg[\prod_{L_j\in\Phi_{\rm u^{'}_l}/u_s}\mathbb{E}_{\rm g_{\rm l}}[\exp(-s\eta_{\rm l}\rho_{\rm u}G_{\rm l}D_{\rm L_j}^{-\alpha_{\rm l}})]\bigg]\nonumber\\
	&\times\mathbb{E}_{\rm \Phi_{\rm c,a}}\bigg[\prod_{C_k\in\Phi_{\rm c,a}\cup C_{\rm R_s}\cup C_{\rm R_s}/u_s}\mathbb{E}_{\rm H}[\exp(-s\rho_{\rm u}H D_{\rm C_k}^{-\alpha_{\rm t}})]\bigg]\nonumber\\
	\stackrel{(a)}{=}& \mathbb{E}_{\rm \Phi_{\rm u^{'}_n}}\bigg[\prod_{N_i\in\Phi_{\rm u^{'}_n}/u_s}\bigg(\frac{m_{\rm n}}{m_{\rm n}+s\eta_{\rm n}\rho_{\rm u}D^{-\alpha_{\rm n}}_{\rm N_i}}\bigg)^{m_{\rm n}}\bigg]\times\mathbb{E}_{\rm \Phi_{\rm u^{'}_l}}\bigg[\prod_{L_j\in\Phi_{\rm u^{'}_l}/u_s}\bigg(\frac{m_{\rm l}}{m_{\rm l}+s\eta_{\rm l}\rho_{\rm u}D^{-\alpha_{\rm l}}_{\rm L_j}}\bigg)^{m_{\rm l}}\bigg]\nonumber\\
	& \times \mathbb{E}_{\rm \Phi_{\rm c,a}}\bigg[\prod_{C_k\in\Phi_{\rm c,a}\cup C_{\rm R_s}/u_s}\bigg(\frac{1}{1+s\rho_{\rm u}D_{\rm C_k}^{-\alpha_{\rm l}}}\bigg)\bigg]\nonumber\\
	\stackrel{(b)}{=}&\exp\biggl(-2\pi\lambda_{\rm u}^{'}\int_{a(\|x\|)}^{\infty}\bigg[1-\bigg(\frac{m_{\rm n}}{m_{\rm n}+s\eta_{\rm n}\rho_{\rm u}(z^2+h^2)^{-\frac{\alpha_{\rm n}}{2}}}\bigg)^{m_{\rm n}}\bigg]zP_{\rm n}(\sqrt{z^2+h^2}){\rm d}z\biggl)\nonumber\\
	&\times \exp\biggl(-2\pi\lambda_{\rm u}^{'}\int_{b(\|x\|)}^{\infty}\bigg[1-\bigg(\frac{m_{\rm l}}{m_{\rm l}+s\eta_{\rm l}\rho_{\rm u}(z^2+h^2)^{-\frac{\alpha_{\rm l}}{2}}}\bigg)^{m_{\rm l}}\bigg]zP_{\rm l}(\sqrt{z^2+h^2}){\rm d}z\biggl)\nonumber\\
	&\times \exp\biggl(-2\pi\lambda_{\rm c}^{'}\int_{c(\|x\|)}^{\infty}\bigg[1-\bigg(\frac{1}{1+s\rho_{\rm u}z^{-\alpha_{\rm t}}}\bigg)\bigg]z{\rm d}z\biggl),
	\end{align}
	where step (a) follows from the moment generating function (MGF) of Gamma distribution, (b) follows from the PGFL of inhomogeneous PPP, $a(\|x\|)$, $b(\|x\|)$ and $c(\|x\|)$ are defined in Lemma \ref{lem_Laplace}.
	
	\subsection{Proof of Theorem~\ref{lem_P_cov}}\label{app_P_cov}
	When the typical user is associated with the LoS UAV in its hotspot center, the coverage probability is given by 
	\begin{small}
		\begin{align}
		\label{pro_eq_P_cov_Uo_L}
		&P_{\rm cov,U_{o,l}} = \mathbb{E}_{\rm R_{\rm u_o}}\bigg[\mathbb{P}\bigg(\frac{\eta_{\rm l}\rho_{\rm u}G_{\rm l}R_{\rm u_{o}}^{-\alpha_{\rm l}}}{\sigma^2+I}\geq\theta|R_{\rm u_{o}}\bigg)P_{\rm l}(R_{\rm u_o})\bigg] = \mathbb{E}_{\rm R_{\rm u_o}}\bigg[\mathbb{P}\bigg(G_{\rm l}\geq \frac{\theta R_{\rm u_o}^{\alpha_{\rm l}}(\sigma^2+I)}{\eta_{\rm l}\rho_{\rm u}}|R_{\rm u_o}\bigg)P_{\rm l}(R_{\rm u_o})\bigg] \nonumber\\
		&\stackrel{(a)}{=} \mathbb{E}_{\rm R_{\rm u_o}}\bigg[\mathbb{E}_{\rm \sigma^2+I}\bigg[\frac{\Gamma_{u}(m_{\rm l},m_{\rm l}g_{\rm l}(R_{\rm u_o})(\sigma^2+I))}{\Gamma(m_{\rm l})}\bigg]P_{\rm l}(R_{\rm u_o})\bigg]\nonumber\\
		&\stackrel{(b)}{=} \mathbb{E}_{\rm R_{\rm u_o}}\bigg[\mathbb{E}_{\rm \sigma^2+I}\bigg[e^{-m_{\rm l}g_{\rm l}(R_{\rm u_o})(\sigma^2+I)}\sum_{k=0}^{m_{\rm l}-1}\frac{(m_{\rm l}g_{\rm l}(R_{\rm u_o})(\sigma^2+I))^{k}}{k!}\bigg]P_{\rm l}(R_{\rm u_o})\bigg]\nonumber\\
		&= \mathbb{E}_{\rm R_{\rm u_o}}\bigg[P_{\rm l}(R_{\rm u_o})\sum_{k=0}^{m_{\rm l}-1}\frac{(m_{\rm l}g_{\rm l}(R_{\rm u_o}))^k}{k!}\mathbb{E}_{\sigma^2+I}\bigg[e^{-m_{\rm l}g_{\rm l}(R_{\rm u_o})(\sigma^2+I)}(\sigma^2+I)^{k}\bigg]\bigg] \nonumber\\
		&\stackrel{(c)}{=} \mathbb{E}_{\rm R_{\rm u_o}}\bigg[\sum_{k=0}^{m_{\rm l}-1}\frac{(-m_{\rm l}g_{\rm l}(R_{\rm u_o}))^k}{k!}\bigg[P_{\rm C,a}\frac{\partial^{k}}{\partial s^{k}}\mathcal{L}_{\rm \sigma^2+I,a}(s,R_{\rm u_o})+(1-P_{\rm C,a})\frac{\partial^{k}}{\partial s^{k}}\mathcal{L}_{\rm\sigma^2+I,n}(s,R_{\rm u_o})\bigg]_{s=m_{\rm l}g_{\rm l}(R_{\rm u_o})}P_{\rm l}(R_{\rm u_o})\bigg]\nonumber\\
		&\stackrel{(d)}{=} \int_{h}^{\sqrt{h^2+r_{c}^{2}}}\sum_{k=0}^{m_{\rm l}-1}\frac{(-m_{\rm l}g_{\rm l}(r))^k}{k!}\frac{\partial^{k}}{\partial s^{k}}\bigg[P_{\rm C,a}\mathcal{L}_{\rm\sigma^2+I,a}(s,r)+(1-P_{\rm C,a})\mathcal{L}_{\rm\sigma^2+I,n}(s,r)\bigg]_{s=m_{\rm l}g_{\rm l}(r)}P_{\rm l}(r) \frac{2r}{r_{c}^2}{\rm d}r,
		\end{align}
	\end{small}
	where $	g_{\rm l}(r)$ and $g_{\rm n}(r)$ are defined in (\ref{eq_g_l_g_n}), step $(a)$ is due to the definition of CCDF of Gamma function $F_{\rm G}(g)=\frac{\Gamma_{u}(m,mg)}{\Gamma(m)}$, with $\Gamma_{u}(m,mg)$ is the upper incomplete Gamma function \cite{8833522}, step $(b)$ follows from $\frac{\Gamma_{u}(m,mg)}{\Gamma(m)}=\exp(-mg)\sum_{k=0}^{m-1}\frac{(mg)^k}{k!}$, step $(c)$ is obtained by  $\mathbb{E}_{U}[\exp(-sU)U^{k}]=(-1)^k\frac{\partial^{k}}{\partial s^{k}}\mathcal{L}_{U}(s)$, and step $(d)$ follows the distribution of $R_{\rm u_o}$ defined in (\ref{dist_u_o}). 
	
	Conditioned on $R_{\rm s}$ and $R_{\rm c}$, and given that the typical UAV is unavailable, the coverage probability when associating with the nearest LoS UAV $P_{\rm cov,U^{'}_l}$ can be written as
	\begin{align}
	\label{pro_eq_P_cov_Up_L}
	P_{\rm cov,U^{'}_l}&= \mathbb{E}_{\rm R_{U^{'},l}}\bigg[\mathcal{A}_{\rm LoS,a}(R_{\rm u^{'},l}|R_{\rm s},R_{\rm c})P_{\rm Crs,a}\mathbb{P}\bigg(\frac{\eta_{\rm l}\rho_{\rm u}G_{\rm l}R_{\rm u^{'},l}^{-\alpha_{\rm l}}}{\sigma^2+I}\geq\theta|R_{\rm u^{'},l}\bigg)\bigg] \nonumber\\
	&+\mathbb{E}_{\rm R_{u^{'},l}}\bigg[\mathcal{A}_{\rm LoS,n}(R_{\rm u^{'},l}|R_{\rm s},R_{\rm c})(1-P_{\rm Crs,a})\mathbb{P}\bigg(\frac{\eta_{\rm l}\rho_{\rm u}G_{\rm l}R_{\rm u^{'},l}^{-\alpha_{\rm l}}}{\sigma^2+I}\geq\theta|R_{\rm u^{'},l}\bigg)\bigg]\nonumber\\
	&= \mathbb{E}_{\rm R_{u^{'},l}}\bigg[\mathcal{A}_{\rm LoS,a}(R_{\rm u^{'},l}|R_{\rm s},R_{\rm c})P_{\rm Crs,a}\sum_{k=0}^{m_{\rm l}-1}\frac{(-m_{\rm l}g_{\rm l}(R_{\rm u^{'},l}))^k}{k!}\frac{\partial^{k}}{\partial s^{k}}\mathcal{L}_{\rm\sigma^2+I,a}(s,R_{\rm u^{'},l})|_{ s=m_{\rm l}g_{\rm l}(R_{\rm u^{'},l})}\bigg]\nonumber\\
	&+\mathbb{E}_{\rm R_{u^{'},l}}\bigg[\mathcal{A}_{\rm LoS,n}(R_{\rm u^{'},l}|R_{\rm s},R_{\rm c})(1-P_{\rm Crs,a})\sum_{k=0}^{m_{\rm l}-1}\frac{(-m_{\rm l}g_{\rm l}(R_{\rm u^{'},l}))^k}{k!}\frac{\partial^{k}}{\partial s^{k}}\mathcal{L}_{\rm\sigma^2+I,n}(s,R_{\rm u^{'},l})|_{s=m_{\rm l}g_{\rm l}(R_{\rm u^{'},l})}\bigg]\nonumber\\
	&= \int_{h}^{\infty}\bigg[\mathcal{A}_{\rm LoS,a}(r|R_{\rm s},R_{\rm c})(1-P_{\rm Crs,a})\sum_{k=0}^{m_{\rm l}-1}\frac{(-m_{\rm l}g_{\rm l}(r))^k}{k!}\frac{\partial^{k}}{\partial s^{k}}\mathcal{L}_{\rm\sigma^2+I,n}(s,r)|_{ s=m_{\rm l}g_{\rm l}(r)}\nonumber\\
	& +\mathcal{A}_{\rm LoS}(r|R_{\rm s},R_{\rm c})P_{\rm Crs,a}\sum_{k=0}^{m_{\rm l}-1}\frac{(-m_{\rm l}g_{\rm l}(r))^k}{k!}\frac{\partial^{k}}{\partial s^{k}}\mathcal{L}_{\rm\sigma^2+I,a}(s,r)|_{s=m_{\rm l}g_{\rm l}(r)}\bigg]f_{\rm R_{u^{'},l}}(r) {\rm d}r.
	\end{align}
	where $f_{\rm R_{u^{'},l}}(r)$ is given in (\ref{dist_u_p_l}). Taking the expectation over $R_{\rm s}$ and $R_{\rm c}$ completes the proof. $P_{\rm cov,U^{'}_n}$, $P_{\rm cov,Cs}$ and $P_{\rm cov,Cc}$ follow a similar method, therefore omitted here.
	
	\subsection{Proof of Lemma~\ref{lem_approximation}}\label{app_approximation}
	Given that (\ref{pro_eq_P_cov_Uo_L}) and (\ref{pro_eq_P_cov_Up_L}) require higher-order derivatives of Laplace transform, we here use the upper bound of the CDF of the Gamma distribution in order to compute a less complicated approximation. It has been shown in both \cite{8833522} and \cite{6932503} that this upper bound  provides a tight approximation to coverage probability, which can be derived as follows
	\begin{align}
	& \mathbb{E}_{\rm\sigma^2+I}\bigg[\frac{\Gamma_{u}(m_{\rm \{l,n\}},m_{\rm l}g_{\rm \{l,n\}}(r)(\sigma^2+I))}{\Gamma(m_{\rm \{l,n\}})}\bigg]= \mathbb{E}_{\rm\sigma^2+I}\bigg[1-\frac{\Gamma_{l}(m_{\rm \{l,n\}},m_{\rm \{l,n\}}g_{\rm \{l,n\}}(r)(\sigma^2+I))}{\Gamma(m_{\rm \{l,n\}})}\bigg]\nonumber\\
	=& 1-\mathbb{E}_{\rm\sigma^2+I}\bigg[\frac{\Gamma_{l}(m_{\rm \{l,n\}},m_{\rm \{l,n\}}g_{\rm \{l,n\}}(r)(\sigma^2+I))}{\Gamma(m_{\rm \{l,n\}})}\bigg]
	\nonumber\\ \stackrel{(a)}{\approx}& 1-\mathbb{E}_{\rm\sigma^2+I}\bigg[\bigg(1-e^{-\beta_{2}(m_{\rm \{l,n\}})m_{\rm \{l,n\}}g_{\rm \{l,n\}}(r)(\sigma^2+I)}\bigg)^{m_{\rm \{l,n\}}}\bigg]\nonumber\\
	\stackrel{(b)}{=}& 1-\mathbb{E}_{\rm\sigma^2+I}\bigg[\sum_{k=0}^{m_{\rm \{l,n\}}}\binom{m_{\rm \{l,n\}}}{k}(-1)^{k} e^{-k\beta_{2}(m_{\rm \{l,n\}})m_{\rm \{l,n\}}g_{\rm \{l,n\}}(r)(\sigma^2+I)}\bigg]
	\nonumber\\
	=& \sum_{k=1}^{m_{\rm \{l,n\}}}\binom{m_{\rm \{l,n\}}}{k}(-1)^{k+1} \mathbb{E}_{\rm\sigma^2+I}\bigg[e^{-k\beta_{2}(m_{\rm \{l,n\}})m_{\rm \{l,n\}}g_{\rm \{l,n\}}(r)(\sigma^2+I)}\bigg]\nonumber\\
	=& \sum_{k=1}^{m_{\rm l}}\binom{m_{\rm \{l,n\}}}{k}(-1)^{k+1}\mathcal{L}_{\rm \sigma^2+I}(k\beta_{2}(m_{\rm \{l,n\}})m_{\rm \{l,n\}}g_{\rm \{l,n\}}(r),r),
	\end{align}
	in which $\Gamma_{l}(m,mg)$ denotes the lower incomplete Gamma function \cite{8833522} and step $(a)$ follows the upper bound in 
	\begin{align}
	\label{eq_approximation}
	(1-e^{-\beta_{1}(m)mg})^{m}<\frac{\Gamma_{l}(m,mg)}{\Gamma(m)}<(1-e^{-\beta_{2}(m)mg})^m,
	\end{align}
	in which,
	\begin{align}
	\beta_{1}(m) = \left\{ 
	\begin{aligned}
	1, \quad \text{if} \quad m>1,\\
	(m!)^{\frac{-1}{m}}, \quad \text{if} \quad m<1, \\
	\end{aligned} \right.\quad
	\beta_{2}(m) = \left\{ 
	\begin{aligned}
	(m!)^{\frac{-1}{m}}, \quad \text{if} \quad m>1,\\
	1, \quad \text{if} \quad m<1. \\
	\end{aligned} \right.
	\end{align}
	Step $(b)$ results from applying Binomial theorem.
	
	In the case of above approximation, $P_{\rm cov,U_{o,l}}$ in (\ref{pro_eq_P_cov_Uo_L}) and $P_{\rm cov,U^{'}_l}$ in (\ref{pro_eq_P_cov_Up_L})  can be rewritten as
	\begin{align}
	P_{\rm cov,U_{o,\{l,n\}}}=& \mathbb{E}_{\rm R_{\rm u_o}}\bigg[\mathbb{E}_{\rm\sigma^2+I}\bigg[\frac{\Gamma_{u}(m_{\rm \{l,n\}},m_{\rm \{l,n\}}g_{\rm \{l,n\}}(R_{\rm u_o})(\sigma^2+I))}{\Gamma(m_{\rm \{l,n\}})}\bigg]P_{\rm \{l,n\}}(R_{\rm u_o})\bigg]\nonumber\\
	=& \mathbb{E}_{\rm R_{\rm u_o}}\bigg[\sum_{k=1}^{m_{\rm \{l,n\}}}\binom{m_{\rm \{l,n\}}}{k}(-1)^{k+1}\bigg(P_{\rm C,a}\mathcal{L}_{\rm \sigma^2+I,a}(k\beta_{2}(m_{\rm \{l,n\}})m_{\rm \{l,n\}}g_{\rm \{l,n\}}(R_{\rm u_o}),R_{\rm u_o})\nonumber\\
	&+(1-P_{\rm C,a})\mathcal{L}_{\rm \sigma^2+I,n}(k\beta_{2}(m_{\rm \{l,n\}})m_{\rm \{l,n\}}g_{\rm \{l,n\}}(R_{\rm u_o}),R_{\rm u_o})\bigg)P_{\rm \{l,n\}}(R_{\rm u_o})\bigg]\nonumber\\
	=& \sum_{k=1}^{m_{\rm \{l,n\}}}\binom{m_{\rm \{l,n\}}}{k}(-1)^{k+1}\mathbb{E}_{\rm R_{\rm u_o}}\bigg[\bigg(P_{\rm C,a}\mathcal{L}_{\rm \sigma^2+I,a}(k\beta_{2}(m_{\rm \{l,n\}})m_{\rm \{l,n\}}g_{\rm \{l,n\}}(R_{\rm u_o}),R_{\rm u_o})\nonumber\\
	&+(1-P_{\rm C,a})\mathcal{L}_{\rm \sigma^2+I,n}(k\beta_{2}(m_{\rm \{l,n\}})m_{\rm \{l,n\}}g_{\rm \{l,n\}}(R_{\rm u_o}),R_{\rm u_o})\bigg)P_{\rm \{l,n\}}(R_{\rm u_o})\bigg]\nonumber\\
	=& \sum_{k=1}^{m_{\rm \{l,n\}}}\binom{m_{\rm \{l,n\}}}{k}(-1)^{k+1}\int_{h}^{\sqrt{h^2+r_{c}^2}}\bigg(P_{\rm C,a}\mathcal{L}_{\rm \sigma^2+I,a}(k\beta_{2}(m_{\rm \{l,n\}})m_{\rm \{l,n\}}g_{\rm \{l,n\}}(r),r)\nonumber\\
	&+(1-P_{\rm C,a})\mathcal{L}_{\rm \sigma^2+I,n}(k\beta_{2}(m_{\rm \{l,n\}})m_{\rm \{l,n\}}g_{\rm \{l,n\}}(r),r)\bigg)P_{\rm \{l,n\}}(r)\frac{2r}{r_c^2}{\rm d}r,\\
	P_{\rm cov,U^{'}_{\{l,n\}}}=& \mathbb{E}_{\rm R_{u^{'},\{l,n\}}}\bigg[\mathcal{A}_{\rm (m_{\rm \{L,NL\}})oS}(R_{\rm u^{'},\{l,n\}})\mathbb{P}\bigg(\frac{\eta_{\rm \{l,n\}}\rho_{\rm u}G_{\rm \{l,n\}}R_{\rm u^{'},\{l,n\}}^{-\alpha_{\rm \{l,n\}}}}{\sigma^2+I}\geq\theta|R_{\rm u^{'},\{l,n\}}\bigg)\bigg] \nonumber\\
	=& \mathbb{E}_{\rm R_{u^{'},\{l,n\}}}\bigg[\mathcal{A}_{\rm LoS}(R_{\rm u^{'},\{l,n\}})\mathbb{E}_{\rm\sigma^2+I}\bigg[\frac{\Gamma_{u}(m_{\rm \{l,n\}},m_{\rm \{l,n\}}g_{\rm \{l,n\}}(R_{\rm u^{'},\{l,n\}})(\sigma^2+I))}{\Gamma(m_{\rm \{l,n\}})}\bigg]\bigg] \nonumber\\
	=& \sum_{k=1}^{m_{\rm \{l,n\}}}\binom{m_{\rm \{l,n\}}}{k}(-1)^{k+1}\mathbb{E}_{\rm R_{u^{'},\{l,n\}}}\bigg[\mathcal{A}_{\rm \{L,NL\}oS}(R_{\rm u^{'},\{l,n\}})\times\nonumber\\
	&\bigg(P_{\rm Crs,a}\mathcal{L}_{\rm \sigma^2+I,a}(k\beta_{2}(m_{\rm \{l,n\}})m_{\rm \{l,n\}}g_{\rm \{l,n\}}(R_{\rm u^{'},\{l,n\}}),R_{\rm u^{'},\{l,n\}})\nonumber\\
	&+(1-P_{\rm Crs,a})\mathcal{L}_{\rm \sigma^2+I,n}(k\beta_{2}(m_{\rm \{l,n\}})m_{\rm \{l,n\}}g_{\rm \{l,n\}}(R_{\rm u^{'},\{l,n\}}),R_{\rm u^{'},\{l,n\}})\bigg)\bigg]\nonumber\\
	=& \sum_{k=1}^{m_{\rm \{l,n\}}}\binom{m_{\rm \{l,n\}}}{k}(-1)^{k+1}\int_{h}^{\infty}\mathcal{A}_{\rm \{L,NL\}oS}(r)f_{\rm R_{u^{'},\{l,n\}}}(r)\times\nonumber\\ &\bigg(P_{\rm Crs,a}\mathcal{L}_{\rm \sigma^2+I,a}(k\beta_{2}(m_{\rm \{l,n\}})m_{\rm \{l,n\}}g_{\rm \{l,n\}}(r),r)\nonumber\\
	&+(1-P_{\rm Crs,a})\mathcal{L}_{\rm \sigma^2+I,n}(k\beta_{2}(m_{\rm \{l,n\}})m_{\rm \{l,n\}}g_{\rm \{l,n\}}(r),r)\bigg){\rm d}r.
	\end{align}
	
	\bibliographystyle{IEEEtran}
	\bibliography{Draft_v0.9.bbl}
\end{document}